\documentclass[twocolumn]{aastex6}
\usepackage{graphicx}
\usepackage{datetime}

\date{\today}
\collaborationName{FIGS Team}

\begin{document}
\title{FIGS -- Faint Infrared Grism Survey: Description and Data Reduction}
\author{Norbert Pirzkal\altaffilmark{1}}
\author{Sangeeta Malhotra\altaffilmark{2,3}}
\author{Russell E. Ryan\altaffilmark{1}}
\author{Barry Rothberg\altaffilmark{4,5}}
\author{Norman Grogin\altaffilmark{1}}
\author{Steven L. Finkelstein\altaffilmark{6}}
\author{Anton M. Koekemoer\altaffilmark{1}}
\author{James Rhoads\altaffilmark{2,3}}
\author{Rebecca L. Larson\altaffilmark{6}}
\author{Lise Christensen\altaffilmark{7}}
\author{Andrea Cimatti\altaffilmark{8,9}}
\author{Ignacio Ferreras\altaffilmark{10}}
\author{Jonathan P. Gardner\altaffilmark{3}}
\author{Caryl Gronwall\altaffilmark{11,12}}
\author{Nimish P. Hathi\altaffilmark{1,13}}
\author{Pascale Hibon\altaffilmark{14}}
\author{Bhavin Joshi\altaffilmark{2}}
\author{Harald Kuntschner\altaffilmark{15}}
\author{Gerhardt R. Meurer\altaffilmark{16}}
\author{Robert W. O'Connell\altaffilmark{17}}
\author{Goeran Oestlin\altaffilmark{18}}
\author{Anna Pasquali\altaffilmark{19}}
\author{John Pharo\altaffilmark{2}}
\author{Amber N. Straughn\altaffilmark{10}}
\author{Jeremy R. Walsh\altaffilmark{15}}
\author{Darach Watson\altaffilmark{7}}
\author{Rogier A. Windhorst\altaffilmark{2}}
\author{Nadia L Zakamska\altaffilmark{20}}
\author{Andrew Zirm\altaffilmark{21}}

\altaffiltext{1}{Space Telescope Science Institute, Baltimore, MD, 21210, USA}
\altaffiltext{2}{Arizona State University, Tempe, AZ, 85287, USA}
\altaffiltext{3}{NASA’s Goddard Space Flight Center, Astrophysics Science Division, Code 660, Greenbelt MD 20771 USA}

\altaffiltext{4}{Large Binocular Telescope Observatory,University of Arizona, AZ, 85721, USA}
\altaffiltext{5}{George Mason University, Fairfax, VA 22030, USA}
\altaffiltext{6}{University of Texas at Austin, Austin, TX, 78712, USA}
\altaffiltext{7}{Dark Cosmology Centre, Niels Bohr Institute, University of Copenhagen, 2100, Denmark} 
\altaffiltext{8}{University of Bologna,
Department of Physics and Astronomy,
Via Gobetti 93/2, I-40129, Bologna, Italy} 
\altaffiltext{9}{INAF - Osservatorio Astrofisico di Arcetri,
Largo E. Fermi 5, I-50125, Firenze, Italy}

\altaffiltext{10}{Mullard Space Science Laboratory, University College London, Holmbury St Mary, Dorking, Surrey RH5 6NT, UK}

\altaffiltext{11}{Department of Astronomy and Astrophysics, The Pennsylvania State University, University Park, PA 16802, USA }
\altaffiltext{12}{ Institute for Gravitation and the Cosmos, The Pennsylvania State University, University Park, PA 16802, USA}
\altaffiltext{13}{Aix Marseille Universit\'e, CNRS, LAM, Marseille, 13388, France}
\altaffiltext{14}{Gemini Observatory, Southern Operations, La Serena, Chile}
\altaffiltext{15}{European Southern Observatory, Garching, 85748, Germany}
\altaffiltext{16}{International Centre for Radio Astronomy Research, The University of Western Australia, Crawley WA 6009, Australia}
\altaffiltext{17}{The University of Virginia, Charlottesville, VA, 22904-4325, USA }
\altaffiltext{18}{Stockholm University, Stockholm, SE-10691, Sweeden}
\altaffiltext{19}{Astronomisches Rechen Institut, Zentrum f{\"ur} Astronomie der
Universit{\"a}t Heidelberg,
D-69120 Heidelberg, Germany}
\altaffiltext{20}{Department of Physics and Astronomy, Johns Hopkins University, Baltimore MD 21218
} 
\altaffiltext{21}{University of Copenhagen, Niels Bohr Institute, København, DK-2100, Denmark}

\begin{abstract}
The Faint Infrared Grism Survey (FIGS) is a deep Hubble Space Telescope (HST) WFC3/IR (Wide Field Camera 3 Infrared) slitless spectroscopic survey of four deep fields. Two fields are located in the Great Observatories Origins Deep Survey-North (GOODS-N) area and two fields are located in the Great Observatories Origins Deep Survey-South (GOODS-S) area. One of the southern fields selected is the Hubble Ultra Deep Field. Each of these four fields were observed using the WFC3/G102 grism (0.8$\mu$m-1.15$\mu$m continuous coverage) with a total exposure time of 40 orbits ($\approx$ 100 kilo-seconds) per field. This reaches a $3 \sigma$\ continuum depth of $\approx 26$\ AB magnitudes and probes emission lines to $\sim$  ${\rm 10^{-17}\ erg\ s^{-1}\ cm^{-2}}$. This paper details the four FIGS fields and the overall observational strategy of the project. A detailed description of the Simulation Based Extraction (SBE) method used to extract and combine over 10000 spectra of over 2000 distinct sources brighter than $m_{F105W}=26.5$\ mag  is provided. High fidelity simulations of the observations is shown to {\it significantly} improve the background subtraction process, the spectral contamination estimates, and the final flux calibration.  This allows for the combination of multiple spectra to produce a final high quality, deep,  1D-spectra for each object in the survey.
\end{abstract}

\section{Introduction}\label{sec:intro}

The study of distant galaxies is dependent on how well one can reliably derive accurate redshifts.  
The most accurate method relies on spectroscopic emission or absorption lines, followed by broad spectroscopic features such as the 4000\AA\ and Lyman breaks. This is particularly important when deploying large surveys to catalogue and discern the properties of objects as a function of cosmic epoch. Absent spectroscopy, photometric methods (photo-z) using color selection, often supplemented by theoretical methods such as fitting spectral energy distribution (SED) templates to the data, can provide rough redshift estimates. The power and relative accuracy of photometric methods are dependent on the sample size used, robust statistical analysis, the quality (and appropriateness) of the input model SEDs or empirical template spectra used, and properly calibrating techniques using spectroscopic data sets.  The smaller the sample size, the less reliable photo-z methods are, and should be treated with extreme caution when used for individual objects (as noted in  \cite{Sawicki1997,Liu1998,pirzkal2013b}). Therefore, spectroscopic follow-up is always required to cull samples of false positives and, for example, should always be the final arbiter in rejecting or supporting claims for the most distant objects detected.

However, spectroscopic observations are not without their own set of complications.  
Extracting object parameters other than redshift alone is extremely expensive in terms of aperture and time required, particularly in probing the earliest epochs of galaxy formation. This is of course because more signal-to-noise (S/N) is required to detect continuum, and even higher S/N is required to detect absorption lines or separate close emission lines, particularly at wavelengths greater than $0.8\mu m$ where telluric sky emission and atmospheric absorption greatly affect observations.

Technical issues, such as proper slit alignment, slit losses, fitting multiple slits on a multi-object mask or multiple fiber placement, size and efficiency of integral field unit spectrographs, flux and wavelength calibration, atmospheric extinction at short wavelengths, and the ever increasing atmospheric absorption and thermal sky emission at longer wavelengths, play a significant role in affecting the viability of spectroscopic surveys.  The more dispersed the light is, the more expensive the survey (i.e. the longer the integration times and larger the collecting surface needed).  A work-around for this is not new or novel. Low-resolution wide-field {\it slitless} spectroscopy for detecting faint targets was first developed over 120 years ago at Lick Observatory with the Slitless Quartz Spectrograph for use on the 36'' Crossley Reflector \citep{palmer03}.  This was later refined for surveying the radial velocities of ``extra-galactic nebulae'' \citep{mayall}.  These surveys focused primarily on UV/Optical ($\lambda$ $<$ 0.5 $\mu$m) low-resolution spectroscopy of emission lines.  Many surveys continued over the decades, e.g. \citet{markarian1967,smith1975,macalpine1977,wasilewski1983}, helping to discover and catalogue quasars and emission line galaxies, as well as the search for young stars and their places of formation, within our own Galaxy, e.g. \citet{Dahm2005}.  Grism surveys can be more efficient, requiring less observing time to reach a given S/N, than their slit/grating counterparts, and can be significantly more robust and reliable than photometric redshift surveys, e.g. \citet{Schmidt1978,Schmidt1986}.
Specifically, slitless spectroscopy does not suffer light loss compared to slit spectroscopy, grism surveys are also significantly more efficient in collecting area, and even with multi object slit masks, or fibers, there is a limit to the number of slits or fibers that can be placed on sky within a field of view.  Slitless grism (whether ground or space-based) observations are not hampered by these limitations. Furthermore, the low-resolution of grism surveys results in a gain in S/N.  This is relevant for all telescope aperture sizes.
Yet, as efficient as these surveys are, the push to higher redshifts, fainter targets, and multiple emission lines to extract physical parameters beyond redshift alone, cannot compete with telluric limitations.  At longer wavelengths the sky-brightness and sky emission lines, along with the limitations of what the Earth's atmosphere absorbs severely limits slitless spectroscopic surveys.  With the launch of HST and the installation of improved instrumentation, grism survey work has seen a renaissance in the last two decades.
This resurgence began with the Near Infrared Camera and Multi-Object Spectrometer (NICMOS), which included surveys such as \cite{mccarthy1999}, followed by the Advanced Camera for Surveys (ACS), which, in addition to pointed surveys, made possible grism surveys parallel to HST primary observations (APPLES: \cite{pasquali03}, GRAPES: \cite{pirzkal04}, and PEARS: \cite{pirzkal09}), 3D-HST: \citet{Momcheva2016}, GLASS: \cite{Treu2015}. With the addition of the Wide Field Camera 3 and its ability
to cover 0.2-1.6 $\mu$m (split over two grisms), wide-field grism observations are cornerstone data products for large surveys such as the WFC3 Infrared Spectroscopic Parallel Survey (WISPS: \cite{atek10}), 3DHST: \citep{Brammer2012}, the Grism-Lens Amplified Survey from Space (GLASS: \cite{schmidt14}), and the Faint Infrared Grism Survey (FIGS: described herein). Moving deep grism surveys from the ground to space has led to a vast improvement in our ability to detect fainter targets, opening a new parameter space for both the most distant objects and to lower mass ranges.  

In this paper, we present the data reduction and spectral extraction for the cycle 22 Treasury program: FIGS (Proposal ID:~13779, PI:~S.~Malhotra). FIGS was awarded 160 orbits with the WFC3/IR instrument to survey four distinct fields with {\em five} Position Angles (PAs) for each field using the G102 grism to a depth of $m_{F105W}\approx$ 26.  {\it FIGS is the deepest HST grism survey to date}. FIGS data are ideal for constraining cosmic re-ionization at $z\!\gtrsim\!7$ through detection of Ly$\alpha$ emission 
\citep[e.g.][]{mr04,stern05,Tilvi16}; probing the star-formation histories for red sequence/blue cloud/green valley galaxies at $z\!\sim\!2$ \citep[e.g.][]{pasquali06,ferreras09,ferreras12,bedregal13,whitaker13}; examining the diversity among emission-line galaxies at $z\!\lesssim\!2$ \citep[e.g.][]{straughn08,pirzkal2013a,atek14}; and providing an unbiased redshift census \citep[e.g.][]{ryan07,Brammer2012}.

This paper is organized as follows: in \S~\ref{sec:survey} the survey motivation and design is explained; in \S~\ref{sec:data} details regarding the data reduction, including object catalogs and spectral extraction are given; in \S~\ref{sec:spectra} the combination of spectra at multiple position angles, S/N calculations are explained, and representative examples for several types of astrophysical sources are provided. Throughout this work, magnitudes are provided in AB units \citep{og}.  


\section{Survey Design} \label{sec:survey}
\begin{figure*}
\center
\hbox{
\hspace{-0.in}
\includegraphics[width=3.5in]{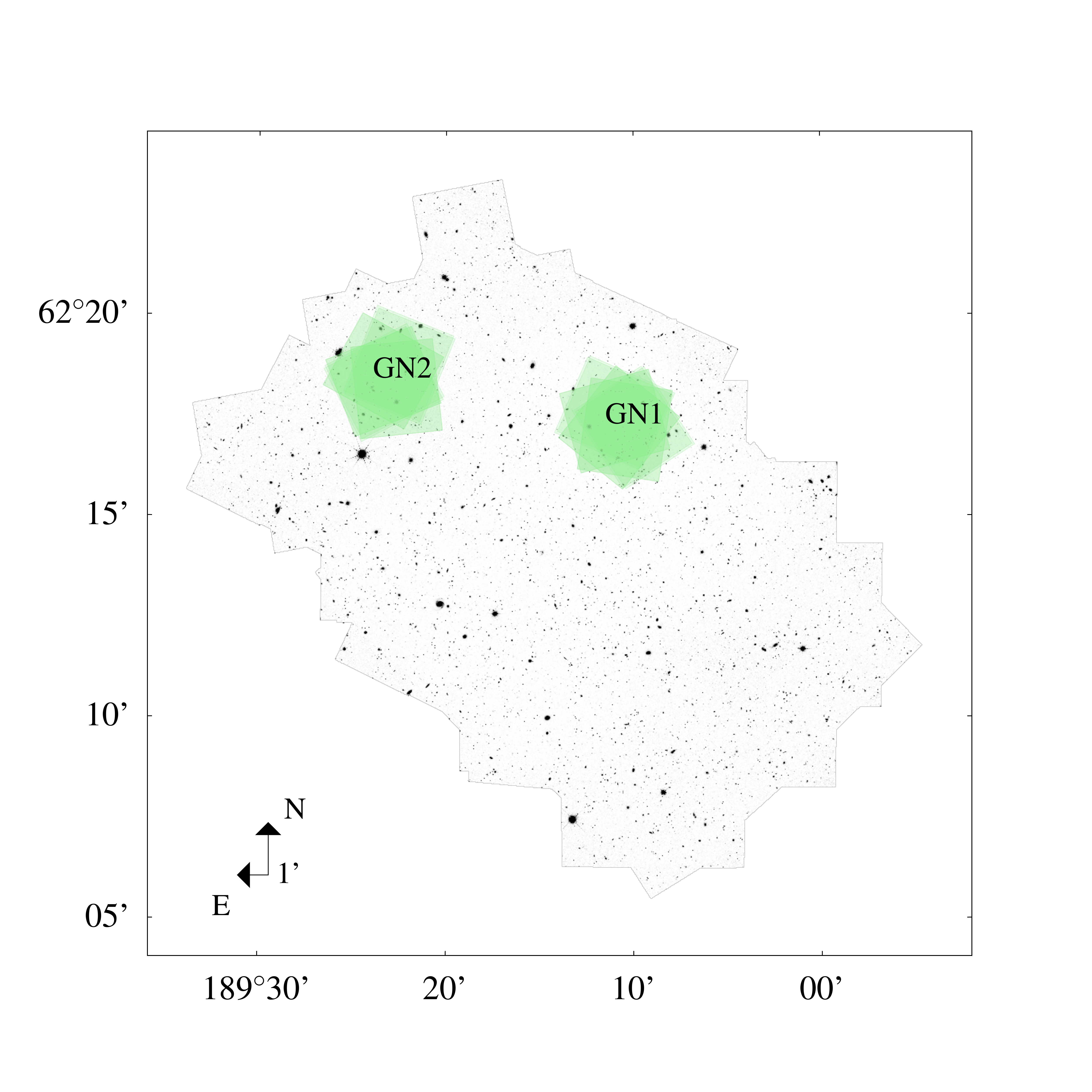}
\hspace{-0.in}
\includegraphics[width=3.5in]{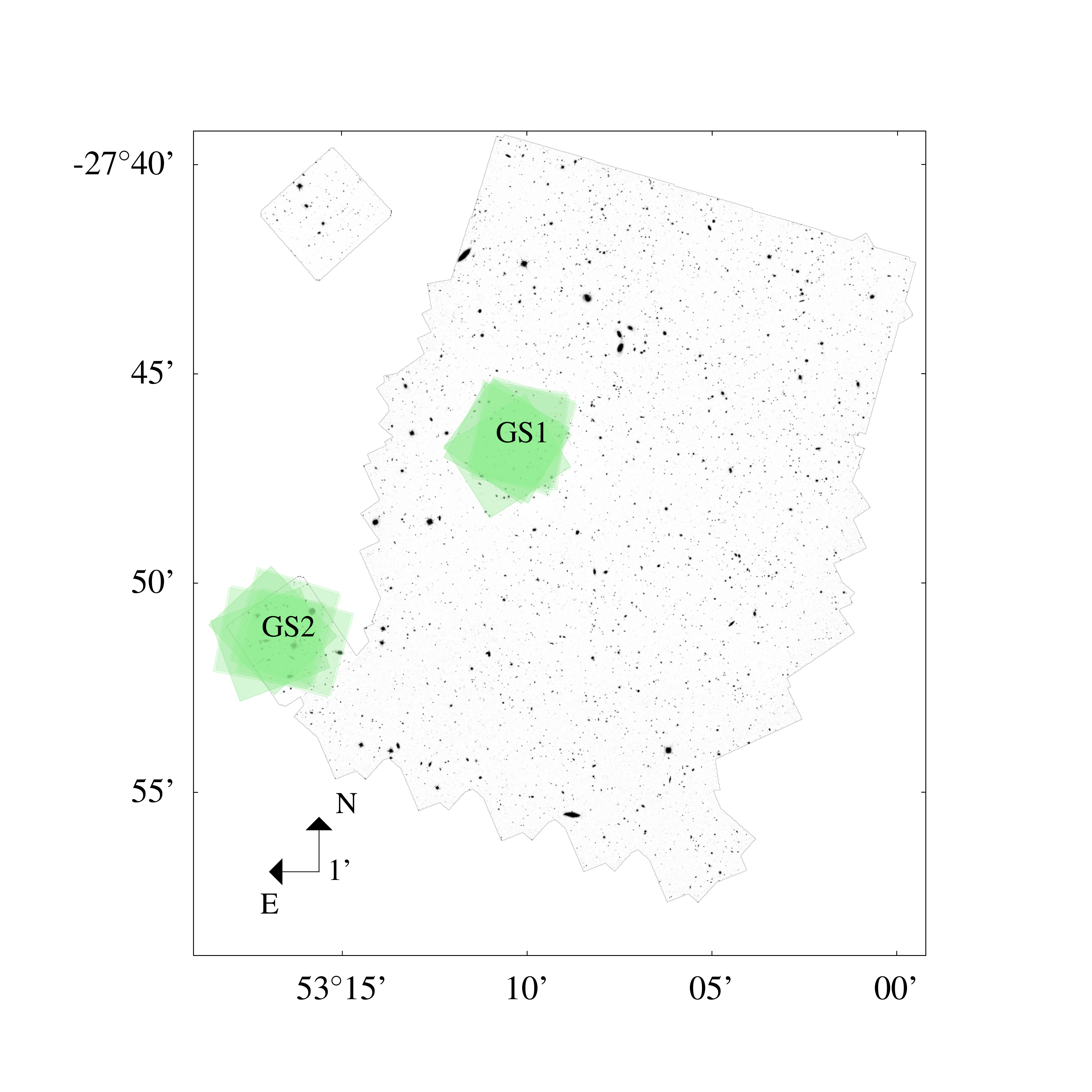}
}

\caption{Left Panel: Location of the FIGS N Fields  (green) with respect to the GOODS-N field. Right Panel: Location of the FIGS S Fields  (green) with respect to the GOODS-S field. The GS1 field is at the same position as the HUDF field. The GS2 field is located at the UDF-PAR2. field. The GOODS-N and GOODS-S mosaics are also shown.  F125W mosaics are shown.}\label{FIGS_NSfields}
\end{figure*}

\begin{figure*}
\center
\hbox{
\hspace{-0.3in}
\includegraphics[width=3.5in]{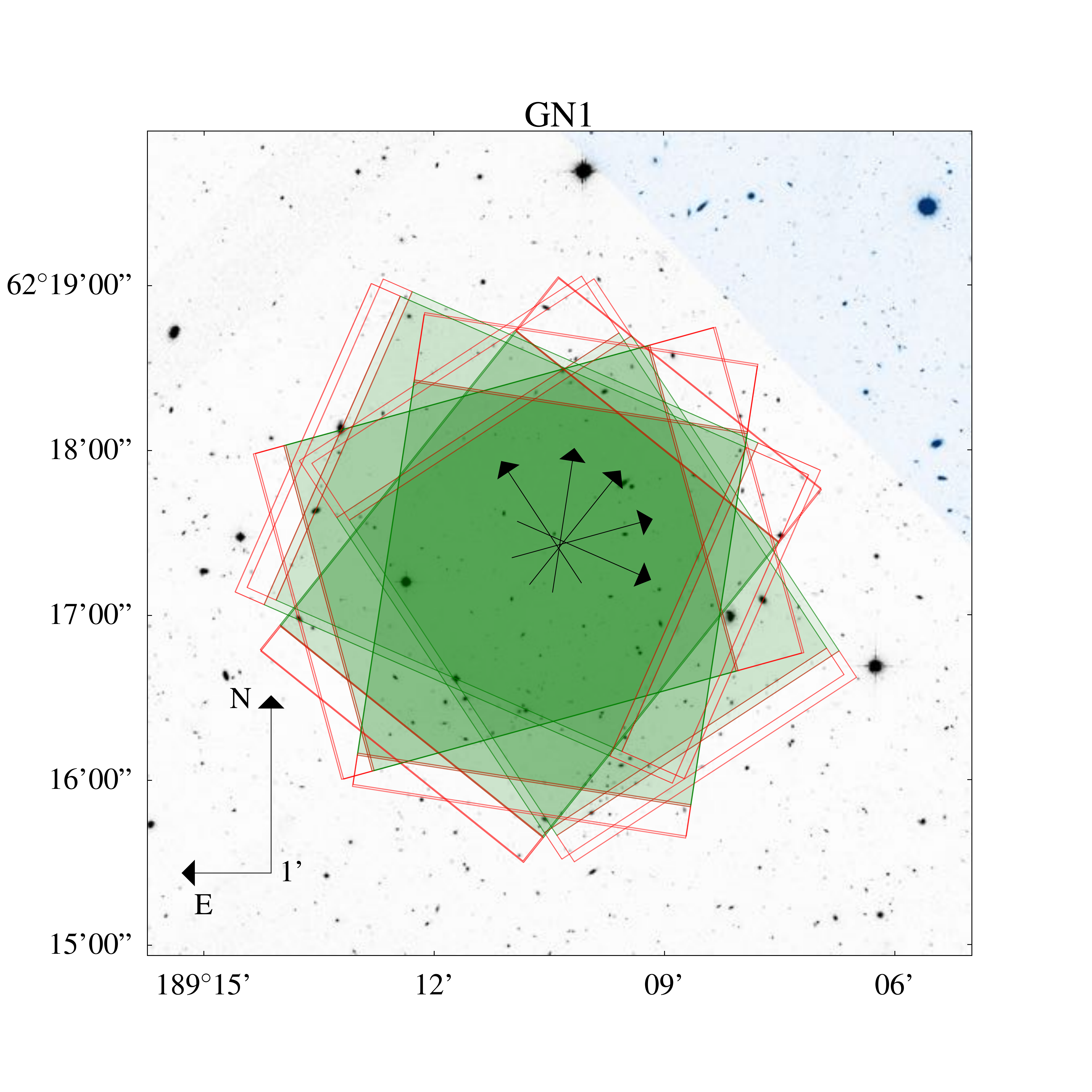}
\hspace{-0.3in}
\includegraphics[width=3.5in]{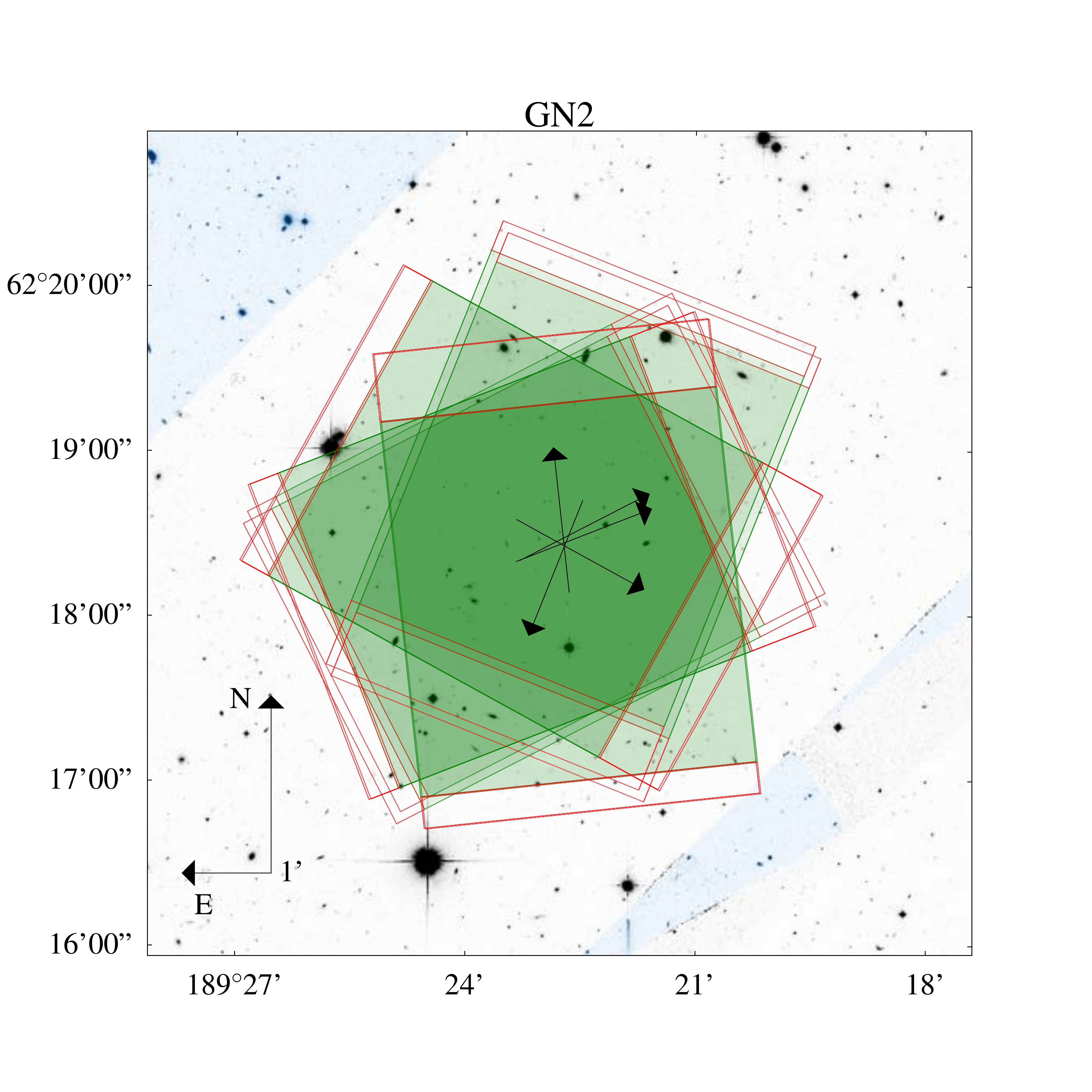}
}
\vspace{-0.3in}
\hbox{
\hspace{-0.3in}
\includegraphics[width=3.5in]{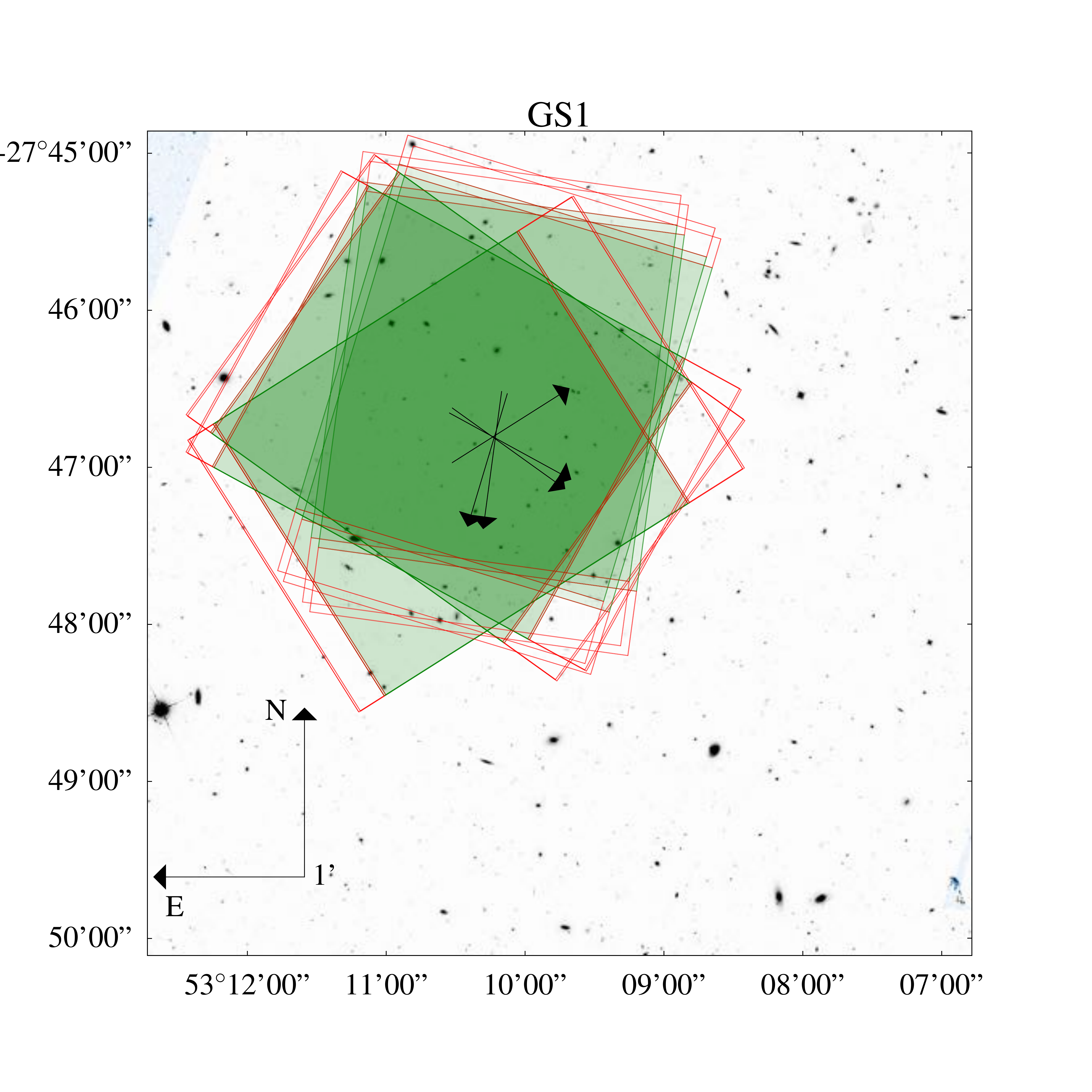}
\hspace{-0.3in}
\includegraphics[width=3.5in]{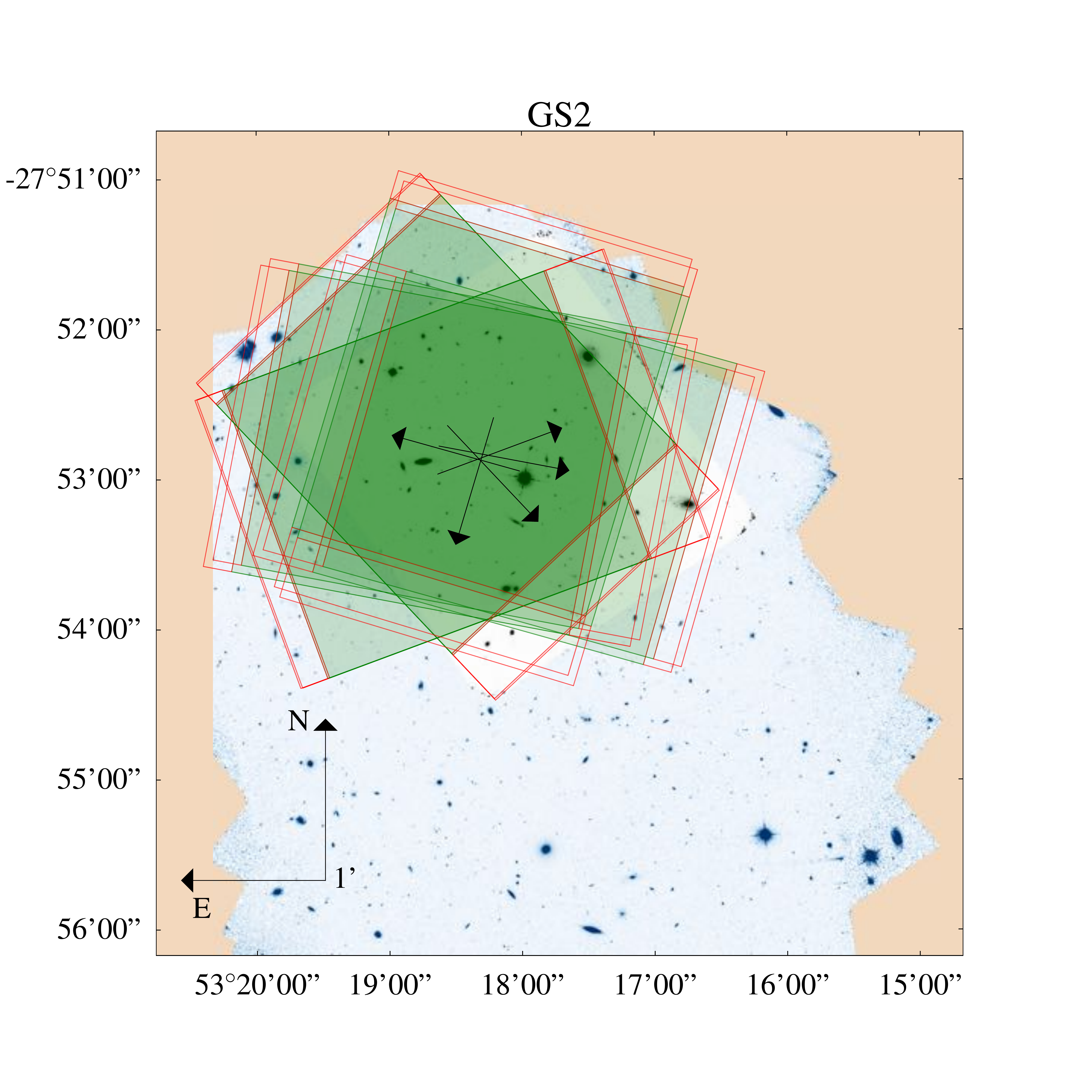}
}
\caption{The four FIGS fields. Each of the four fields were observed at 5 different position angles, as shown by the black arrows at the center of each field.  The regions covered by the grism is shown in green, while regions near the field of view where objects outside of the field of view could still result in dispersed orders are delineated using red lines. The grey mosaic shows areas with existing F105W imaging. Areas with only ACS z-band data and no F105W data are shown in blue. Part of the fields where no imaging was available to determine the existence of contaminating objects are shown in orange. In these plots, North is up and the footprints of individual G102 observations are shown in green and can be directly compared to Figure \ref{FIGS_NSfields} }\label{FIGS_fields}
\end{figure*}

FIGS is designed to maximize coverage and depth of field, while striving to reduce contamination and spurious detections by leveraging multiple PAs.  The fields selected were based upon already available photometric data. Four pointings were selected for the FIGS observations, two in the GOODS-North
region and two in the GOODS-South region (Table \ref{Table1} and Figures \ref{FIGS_NSfields} and \ref{FIGS_fields}).  All four were chosen to
maximize the number of high-redshift ($z>6$) candidates within the WFC3/IR grism
field of view (69, 21, 144, 38 candidates in GN1, GN2, GS1, and GS2, respectively).  These candidates were selected via SED fitting to the available HST deep multi-filter broadband imaging from $B$-band through to $H$-band \citep{Finkelstein2015}.
These broadband exposures came from several different programs, including GOODS \citep{Giavalisco2004}, CANDELS \citep{Grogin2011,Koekemoer2011}, and successive HUDF campaigns \citep{Beckwith2006}.  In the GOODS-South region,
one pointing was situated within the HUDF, and the other was situated within
a HUDF parallel field, slightly displaced from the GOODS area.  The two
GOODS-North fields were both within the CANDELS deep near-infrared area.

\begin{figure*}
\center
\includegraphics[width=3.5in]{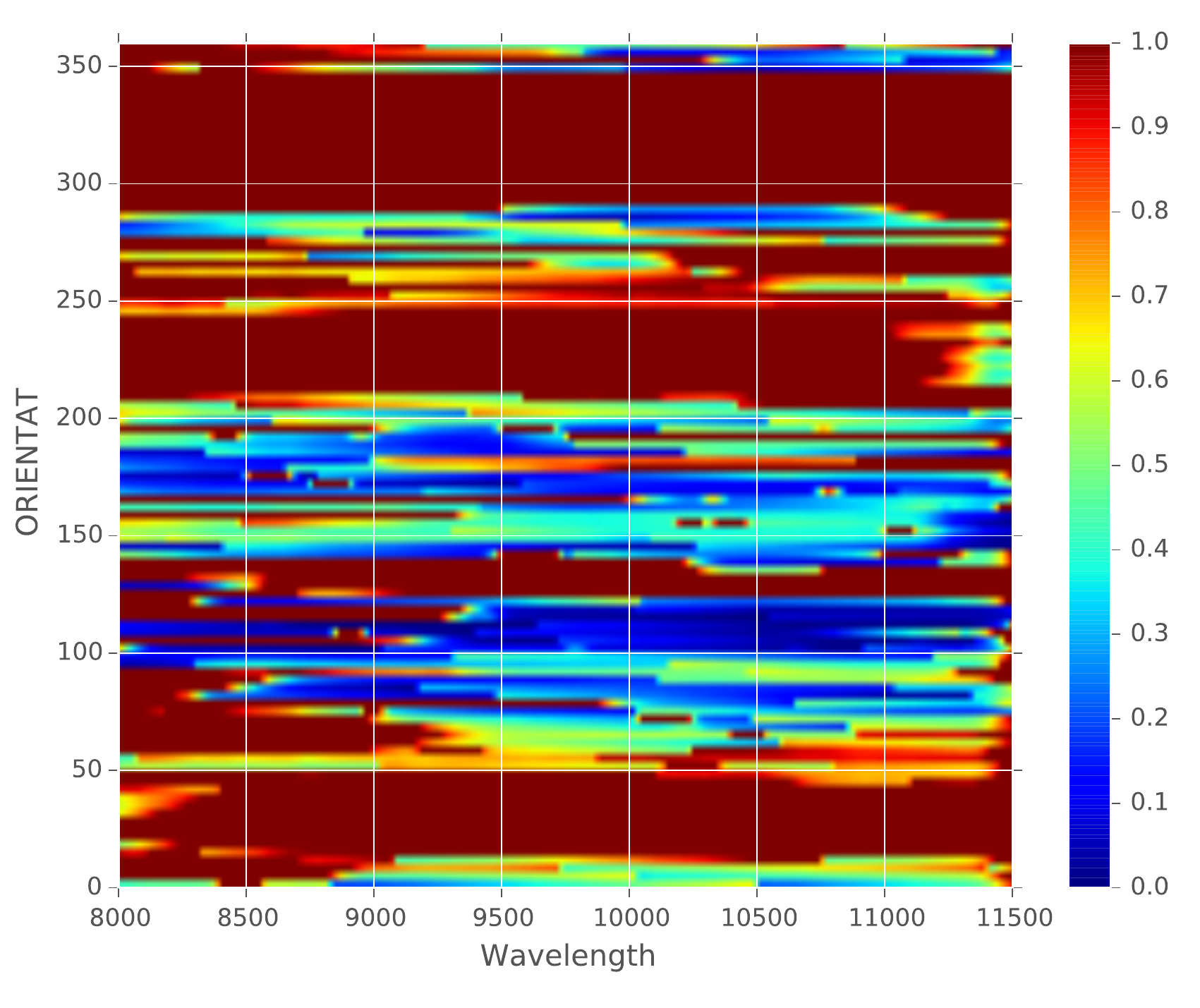}
\caption{Example of one of the diagnostic plots used to determine the position angles (ORIENTAT) where a source of interest was the least contaminated. Regions of high contamination are shown in red while low contamination regions appear in blue. In this example, the lowest amount of contamination occurs at a position angle of 99 degrees. This plot shows the fraction of the observed flux that is due to contamination as a function of position angle and wavelength. Plots such as this one were generated for each of our high-redshift ($z>6$) candidates. We then manually selected specific position angles for each field by manually maximizing the number of some of the brighter high-z targets likely to be uncontaminated. \label{Orient}}
\end{figure*}

For each pointing location, the FIGS team simulated WFC3/IR grism exposures as a 
function of HST roll angle, using the existing near-infrared images. We generated simulations with orientations ranging between 0 and 360 degrees in steps of 0.5 degrees. At each orientation the ratio of contaminating flux to source flux was computed where the first order spectrum of a particular high redshift candidate was located. Plots such as the one shown in Figure \ref{Orient} were generated to determine position angles where contamination was as low as possible (shown in blue). The
simulations were then checked for dispersed source overlap, to identify five 
roll angles in each field that minimized grism contamination of the high-redshift candidates.  The selected roll angles for each field are listed in Table \ref{Table1}. Some of these roll angles are close together but we ensured that at least 3 significantly different roll angles were obtained for each field.
Four 2-orbit visits were obtained for each roll angle with the  G102 grism, resulting in 40 orbits per FIGS field and
approximately 100 ksec of exposure time per field.

The sequencing of grism and direct imaging exposures within each 2-orbit HST visit was carefully tailored to protect the
G102 exposures from illumination by the sunlit Earth limb.  WFC3/IR observations at low limb-angle to the sunlit Earth are known to suffer from much higher and rapidly variable background, and are particularly sensitive to the
air-glow of helium $\lambda10830$\AA\ \citep{Brammer2015}.  
Once the observing window of a FIGS visit was fixed, we determined whether a
given HST visit would be rising or be setting over the sunlit Earth
limb. This information was provided to us by our Program Coordinator (PC) at the Space Telescope Science Institute.  We then placed the broadband F105W alignment exposure either before or after the G102 exposures, as appropriate, to take the brunt of the high-background portion of the orbit.

Because FIGS target fields are each a single WFC3/IR field
of view, we opted for a minimally distributed dither
pattern among observations within a single epoch.  For each
2-orbit visit, we initially chose the IR-DITHER-BLOB
pattern (See Section C.2 of the WFC3 Instrument Handbook) of $\approx3''$ between orbits, with
small intra-orbit dithering ($\approx0.3''$) for improved subpixel-phase
sampling.  The IR-DITHER-BLOB pattern has the added benefit
of displacing the ACS/WFC parallel exposures by slightly
more than the gap between the two ACS/WFC CCDs.  Partway through
the program's execution, we expanded slightly the inter-orbit
dither pattern to match that adopted by the 3D-HST program \citep{Brammer2012},
thereby better mitigating WFC3/IR self-persistence.

The portion of the WFC3/IR field-of-view for which the full G102
trace is available for all targets (as unvignetted by the
detector edge) is not centered on the detector.  To maximize the
number of targets with full grism traces available in all five
epochs, taken at varying HST roll angles, we further introduced an
epoch-dependent dither offset ($\approx17''$) tailored to co-locate the
full-trace region on the sky, regardless of HST orientation.

To maximize S/N in the stack of dithered WFC3/IR
grism exposures, the FIGS team initially chose a 100-second
IR sampling sequence (''SPARS100") with 12--15 samples
per exposure (Exposure times between 1100 and 1400s).  Partway through the execution of the program,
the team concluded that detector self-persistence was sufficiently
severe (i.e. spectra of bright stars left a visible imprint in subsequent exposures at levels greater than 0.005 $e^-/s$) to merit a change to a 50-second
sampling pattern (''SPARS50")  with 12--15 samples per exposures (Exposure times between 450 and 700s).  Although this doubled the number
of exposures and resulted in a modest loss of cumulative IR exposure
time (i.e. increased overhead), the overall sensitivity was improved by the reduction of
self-persistence within the image sequence, avoiding fake/ghost objects and spectra on subsequent, dithered exposures.  The much shorter
exposures for broadband (F105W) alignment used a 25-second
sampling sequence (''SPARS25") with 11--13 samples per exposure.

\begin{deluxetable*}{ccccccc}
\tablecolumns{6} 
 \tablehead{\colhead{Field} & \colhead{RA} & \colhead{DEC} & \colhead{Position} &  \colhead{Number} &  \colhead{Total } \\
 \colhead{Name} & & & \colhead{Angles (PA)} & \colhead{of exposures} &  \colhead{Exposure Time}
 }
\startdata
GN1 & 12h36m42.56s & 62d17m16.89s& --164,--128,--98,--56,156 & 320 & 101120 \\
GN2 &  12h37m32.04s  &  62d18m26.06s & --158,--152,--83,68,151 & 288 & 103823 \\
GS1 & 03h32m41.56s & --27d46m38.80s & --147,73,82,144,151 & 320 & 95469 \\
GS2 & 03h33m06.76s & --27d51m16.56s & --159,--15,73,133,169 & 288 &  98822 \\
\enddata
\caption{Coordinates and exposure time of the four FIGS fields. \label{Table1}}
\end{deluxetable*}

\section{Data Reduction} \label{sec:data}
\subsection{Overview}

The FIGS G102 data were reduced in a manner that loosely follows the method used for the GRAPES and PEARS data \citep{pirzkal04}. However, the planning and execution of the observations were first based on accurate simulations, which then served as the
the basis for the actual extraction of the data.  Simulation Based Extraction (SBE) is a critical step to significantly improving source extraction.   It is essential to first simulate the data so that the astrometric solution and the input object catalogs are validated. Furthermore, access to high fidelity simulations (\S~\ref{simulations}) allows for fine-tuned background subtraction corrections (\S~\ref{back_sub}),  detailed contamination estimates, as well as the appropriate application of object specific sensitivity functions to each extracted spectrum (\S~\ref{sec:combined}).

\subsection{Pre-processing}

\subsubsection{Field Mosaics}\label{mosaics}
For this project we used the full-depth HST ACS and WFC3 mosaics of these fields, predominantly from CANDELS \citep{Grogin2011,Koekemoer2011}, and GOODS \citep{Giavalisco2004}, supplemented as needed by imaging from other HST programs \citep[including the HUDF;][]{Beckwith2006, Oesch2010, Ellis2013, Koekemoer2013, Illingworth2013}. The filters used for detection and target selection in this project were primarily the F850LP in ACS (roughly equivalent to SDSS {\it z}), together with the F125W and F160W in WFC3 (roughly equivalent to {\it J} and {\it H}-band, respectively). These mosaics have pixel scales of 30mas/pixel for the F850LP and 60mas/pixel for the F125W and F160W,and were used as the absolute astrometric reference for the new HST grism imaging that were obtained for these fields.  They also served as detection images for all the spectroscopically detected sources. Further details are provided in \cite{Koekemoer2011} on the image combination and processing approaches that were used to produce these mosaics.

\subsubsection{Object catalogs}\label{catalogs}
We created photometric catalogs using a custom version of Source Extractor \citep{Bertin1996}.  Our modified version adds a buffer between the source and the local background cell, and removes spurious sources associated with the distant wings of bright objects. Catalogs were generated independently in each of our four sub-fields, using a 10000 x 10000 pixel mosaic (30mas per pixel and a size of 5 by 5 arcmin) centered on our grism pointings (see \S \ref{mosaics} for details on image reduction).  We used Source Extractor in two-image mode, where the same detection image (F125W) was used to measure photometry from all available HST filters.  The choice of the F125W was predicated on the fact that the F125W coverage of the FIGS fields is more complete and uniform than the F105W coverage. The choice of the F125W is also more appropriate to detect faint $z>8$\ galaxies since these objects should drop out of F105W images. While, in theory we might expect that some very low continuum galaxies with bright emission lines might be missing from our object extraction catalog, supplemental methods can be used to search for such sources \citep{straughn08,pirzkal2013a,Pirzkal2017c}.

The filters used to provide supplemental photometric information to the F125W detection catalog were: F435W, F606W, F775W and F850LP from the GOODS ACS survey, and F105W, F160W from the CANDELS WFC3/IR survey.  Deep ACS/F814W imaging taken in parallel to CANDELS was also used, as well as the shallow WFC3/IR F140W pre-imaging from 3DHST.  In the GS1 field, which overlaps with the HUDF, we also made use of imaging from the HUDF, HUDF09 (Oesch et al.\ 2010) and UDF12 (Ellis et al.\ 2013).

As this catalog specifies the size and position of sources for extraction from the spectroscopic grism frames, the fidelity of the sources are important.  For this reason, we elected to compose a combined photometric catalog, using both a ``cold" catalog with conservative detection parameters which keep large objects together, and a ``hot" catalog with more aggressive detection parameters to ensure that we include faint sources.  Table \ref{SeXTable} lists the main extraction parameters used for both the cold and hot catalogs.

\begin{table*}
\begin{center}
\begin{tabular}{r|c|c|c|}
\cline{2-3}
\hline
\multicolumn{1}{|c|}{SeXtractor} & \multicolumn{3}{|c|}{Catalog} \\
\cline{2-4}
\multicolumn{1}{|c|}{Parameter} & Cold & Hot & Super-Hot\\
  \hline
\multicolumn{1}{|c|}{DETECT\_THRESH} & 1.5 & 0.7 & 0.35\\
\hline
\multicolumn{1}{|c|}{DETECT\_MINAREA} & 28 & 28 & 14 \\
\hline
\multicolumn{1}{|c|}{DEBLEND\_NTHRESH} & 8 & 32 & 32 \\
\hline
\multicolumn{1}{|c|}{DEBLEND\_MINCONT} & 0.01 & 0.0001 & 0.0001\\
\hline
\multicolumn{1}{|c|}{Filter (9x9 pixel)} &  top-hat & Gaussian & Gaussian \\
    \hline

\end{tabular}
\caption{FIGS SeXtractor Parameters used to generate the cold and hot detection catalogs. \label{SeXTable}}

\end{center}
\end{table*}

A minimum footprint size of 28 pixels (approximately equal to the number of pixels in the point-spread function at the redder WFC3/IR wavelengths) was used for Source Extractor.  The detection and deblending parameters were tuned by inspecting both catalogs, and ensuring that large galaxies remained a single object in the cold catalog, and very faint, yet likely real, objects were still detected in the hot catalog.  Similar to the catalogs from \citet{Finkelstein2010, Finkelstein2012, Finkelstein2015}, we measure colors in small, elliptical Kron \citep{kron80} apertures with PHOT\_AUTOPARAMS set to 1.2, 1.7, and aperture corrections were derived in the F160W-band using the default MAG\_AUTO parameters of 2.5, 3.5 (which has been found to reliably return the total flux within $\sim$5\%).  The default values of 2.5, 3.5 correspond to an aperture which measures the total magnitude to within 6\% accuracy.  However, this aperture does not calculate colors with the optimal S/N, as it includes many sky pixels (a necessary trade-off to accurately measure the total flux).  \citet{Finkelstein2012} showed that using smaller ellipses with PHOT\_AUTOPARAMS set to 1.2, 1.7 more accurately recovered colors for simulated sources  (see also \cite{Bouwens2007}). We therefore measure colors in these smaller apertures, but derive an aperture correction to total in the F160W-band as the ratio of the flux in the larger-to-smaller Kron apertures.

These aperture corrections can fail for objects near to very bright sources, as their Kron radii may not be reliably derived.  In rare cases where aperture correction had non physical negative values, visual inspection showed that these objects were close to brighter objects. For these circumstance, we applied statistical aperture corrections taken as the median aperture corrections for sources of similar fluxes ($\pm$10\%).

This process resulted in a complete hot and cold catalog, containing fluxes for all objects in each catalog in each filter.  These catalogs were then merged into a final catalog.  However, objects in the hot catalog were  only included if their central pixel had a value of zero in the \emph{cold} catalog segmentation map (i.e., it did not lie within the isophotal region of a cold catalog object).  For this comparison, the cold catalog segmentation map was dilated to slightly increase the area of each object (this was only applied during the merging process, and did not affect the photometry).  The combined catalogs were then visually inspected to identify and remove potential non-real objects, such as diffraction spikes from bright objects, or noise spikes picked up in the hot catalog.  We also visually identified objects which were split by the catalog process, and merged them in the final catalog (recalculating the shape-parameters from the final, combined objects).  Faint sources near bright objects were assumed to be part of the bright object, although we performed intensive visual inspection of our catalogs when deriving our extraction parameters to  ensure that we did not under or over-split bright, extended objects. We compared this catalog to the known positions of high redshift galaxies using a master catalog of known objects from \citet{Finkelstein2015} and \citet{Bouwens2015}.  Some of the fainter objects were not present in our catalog.  These objects are desirable, as detecting Ly$\alpha$ emission from such sources is one of the main goals of this survey.  We thus ran a ``super-hot" catalog with DETECT\_THRESH=0.35 and DETECT\_MINAREA=14, which identified missing sources from those catalogs, and added them to our final catalog.

\subsubsection{Astrometric correction}
\begin{figure*}

\includegraphics[width=6.in]{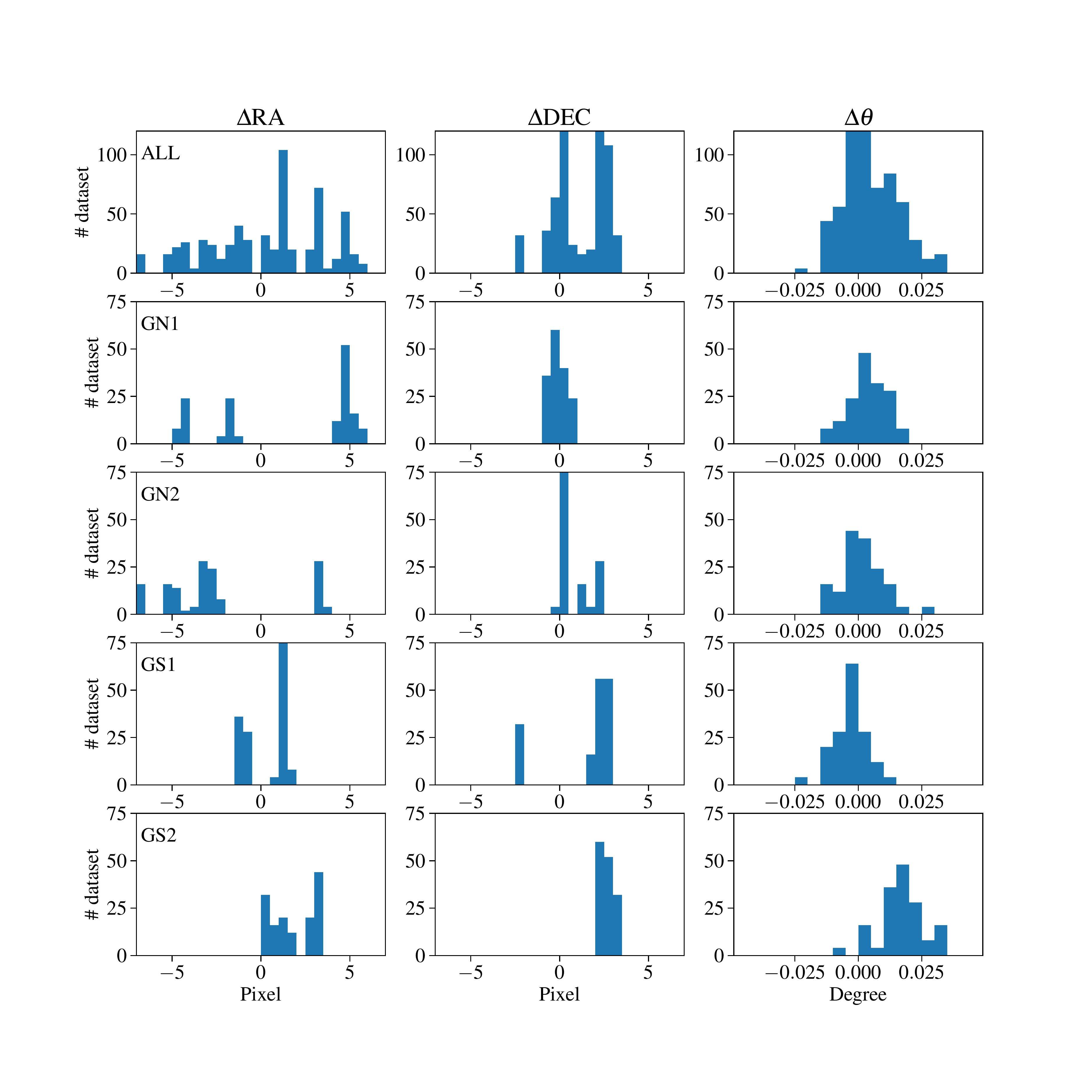}
\caption{Histograms of the corrections applied to the original world coordinate system of the G102 data. We show the full sample of WCS corrections in the top row and individually for the 4 FIGS fields in the following rows. The corrections were $0.4 \pm 3.2$\ pixel in R.A. and $1.17\pm 1.5$\ pixels in DEC, or  $0.05 \pm 0.4$'' and  $0.15 \pm 0.19$'', respectively. Rotations were within $\approx0.03$\ degree. \label{FIGS_offsets}}

\end{figure*}

The astrometric reference frame for the FIGS fields were provided by the large 30mas scale mosaics of the fields discussed in Section \ref{mosaics}. We started by using the SWarp program to generate deep images of the field, properly oriented and with the native WFC3 pixel scale for each of the FIGS visits. Individual FIGS F105W direct images, which were taken during each of the FIGS visit to provide an astrometric reference frame for the G102 exposures, were then astrometrically registered to the deep, rotated mosaics. While we initially used the Astrodrizzle task {\tt Tweakreg} to perform this task, we found that we could not properly control which objects in the field were used to compute the x and y offsets between our FIGS F105W images and our master mosaic images. Unfortunately, the master mosaics were generated with data that are several years old, and each of the FIGS field contains bright stars with a significant amount of proper motion \citep[${\rm < 3mas\ \pm 5 /yr, on\ average;}$][]{windhorst2011}. It was therefore preferable to use many faint sources (galaxy or stars) to astrometrically register these images to our master mosaics, as was done in \cite{Pirzkal2005}. An iterative version of the {\tt FOCAS} \citep{Valdes1995} triangulation algorithm was implemented to register the geometrically distorted FIGS F105W {\it FLT} (HST pipeline calibrated exposure) images to the deep, rectified FIGS mosaics. During each iteration, between 40 and 100 sources were matched and average RA and DEC shifts as well as any needed rotation. These were applied as a correction to the FIGS F105W image and the G102 images taken during the same HST visit. We found this approach to be accurate with residuals on the order of 0.2 WFC3 IR pixels (25mas), or, about 2 times {\it better} than those we were able to achieve previously. Figure \ref{FIGS_offsets} shows histograms of the computed RA and DEC offsets, as well as the rotations needed to match the FIGS data to the reference mosaics, described in Section \ref{mosaics}. As this Figure demonstrates, the computed corrections vary from field to field as different guide stars are used for each field and orientation. The multi-nodal nature of the distribution of the applied offsets are caused by errors in the assumed positions of the HST guide stars. When revisiting the same field, using the same guide stars and orients, this error is approximately 50-100 mas. However, when different guide stars are used, as it is the case when we observed a FIGS field at a different position angle, the error much larger and is typically between 0.2 and 0.5 arcseconds.

\subsubsection{Simulations and Simulations Based Extraction}\label{simulations}
\begin{figure*}
\center
\includegraphics[width=7.in]{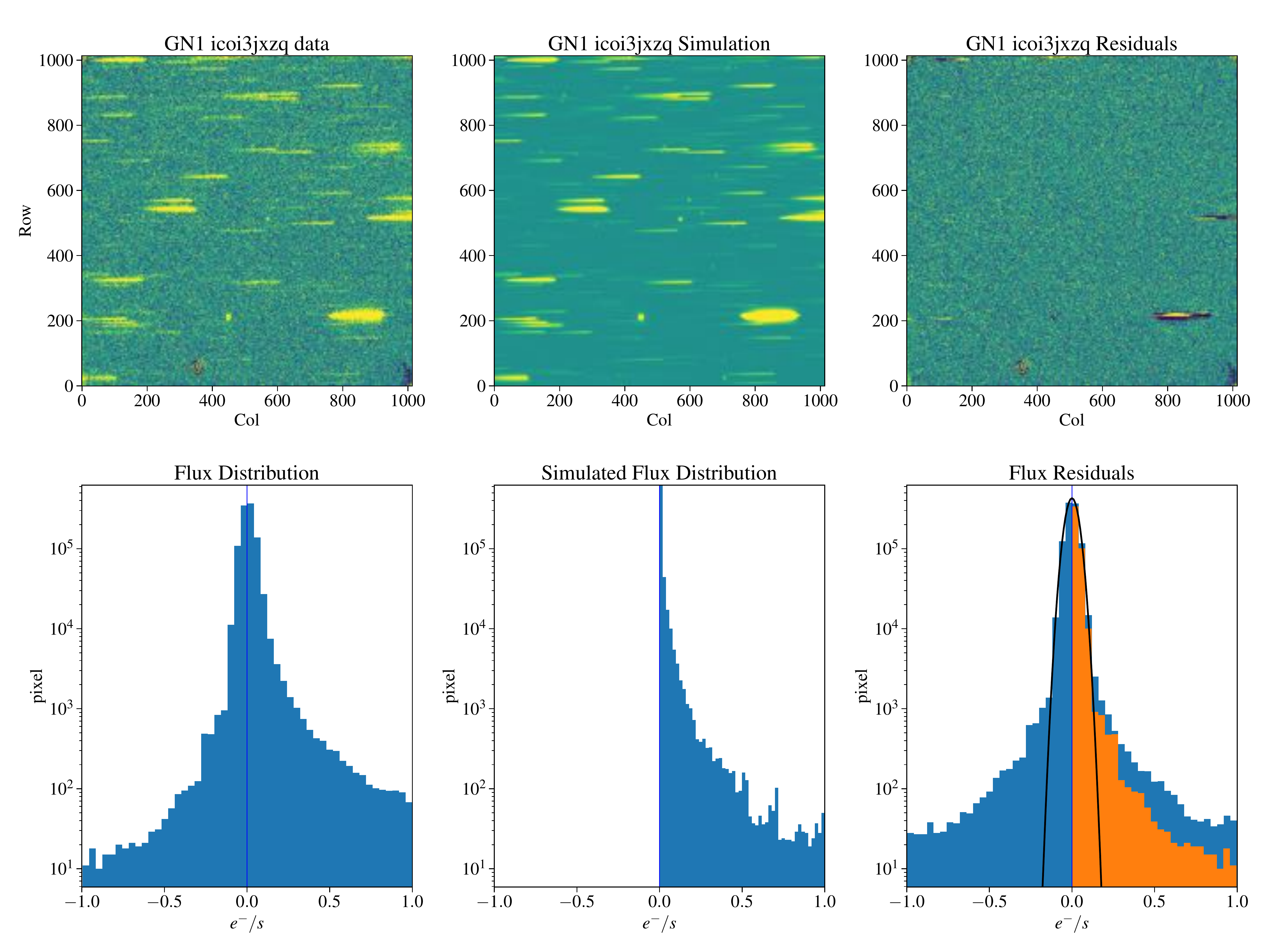}
\caption{We show one of the FIGS dataset for the GN1 field (background subtracted as discussed in Section \ref{back_sub}), the FIGS simulation of the same dataset, and the residuals after subtracting the simulated data from the observed data. A clean subtraction of the dispersed spectra from the FIGS observations (right panel) demonstrates the accuracy of the astrometry, the dispersed background subtraction. It also demonstrates the accuracy of the G102 grism calibration and of the input FIGS photometry which we relied upon to generate our simulations. The bottom right panel shows the expected distribution of Gaussian noise (black line). We over-plotted the negative residual on top of the position residual (orange) to qualitatively show the asymmetry of the histogram of the background residuals. These are expected because of the persistence effect, saturated sources, as well as faint sources possibly missing from our catalog and simulated images. Stars with large proper motion produce regions with larger positive and negative residuals in the top right panel. \label{FIGS_simul_GN1}}
\end{figure*}

\begin{figure*}
\center
\includegraphics[width=7.in]{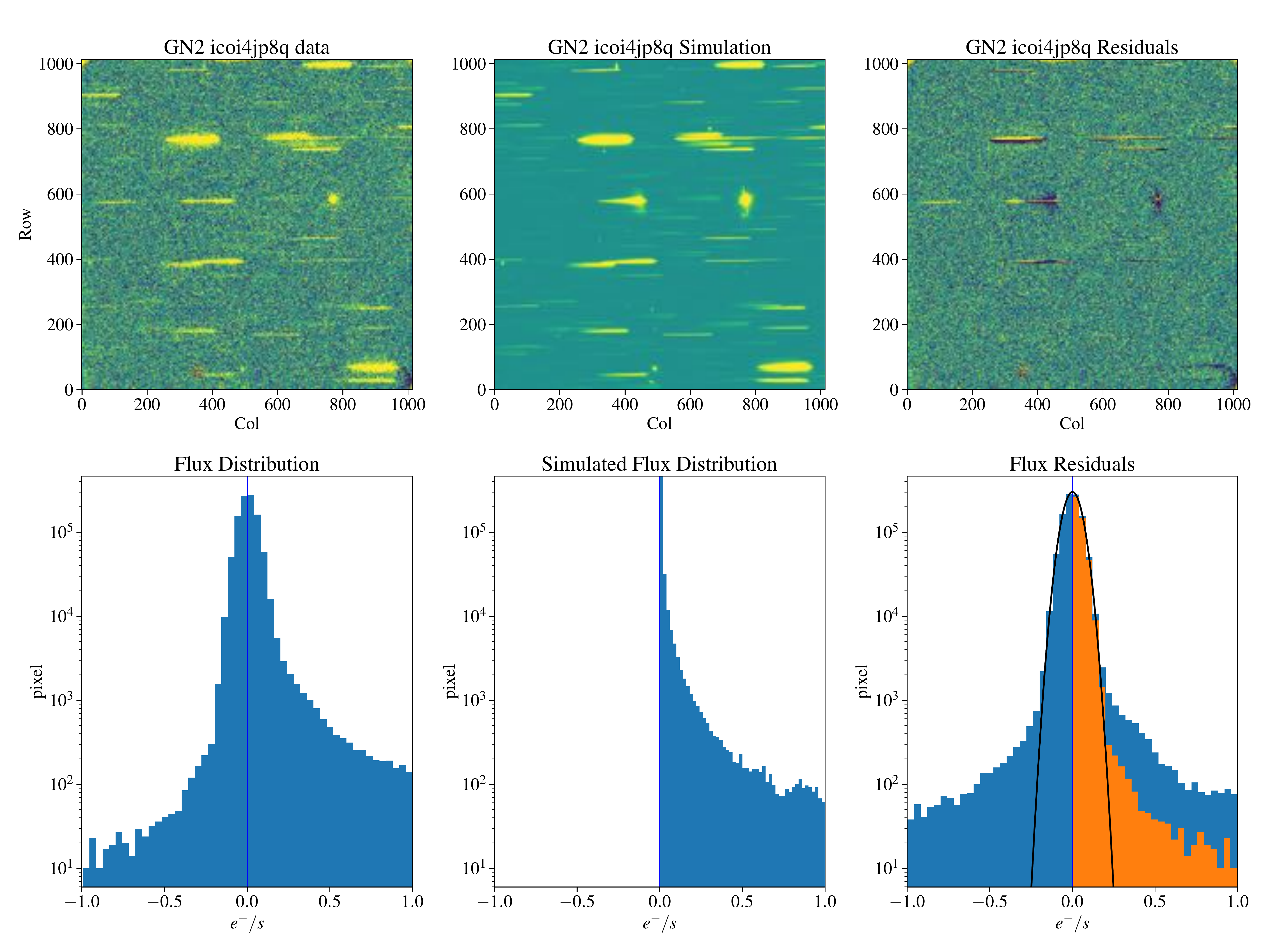}
\caption{Same as Figure \ref{FIGS_simul_GN1} but for the GN2 field. \label{FIGS_simul_GN2}}
\end{figure*}

\begin{figure*}
\center
\includegraphics[width=7.in]{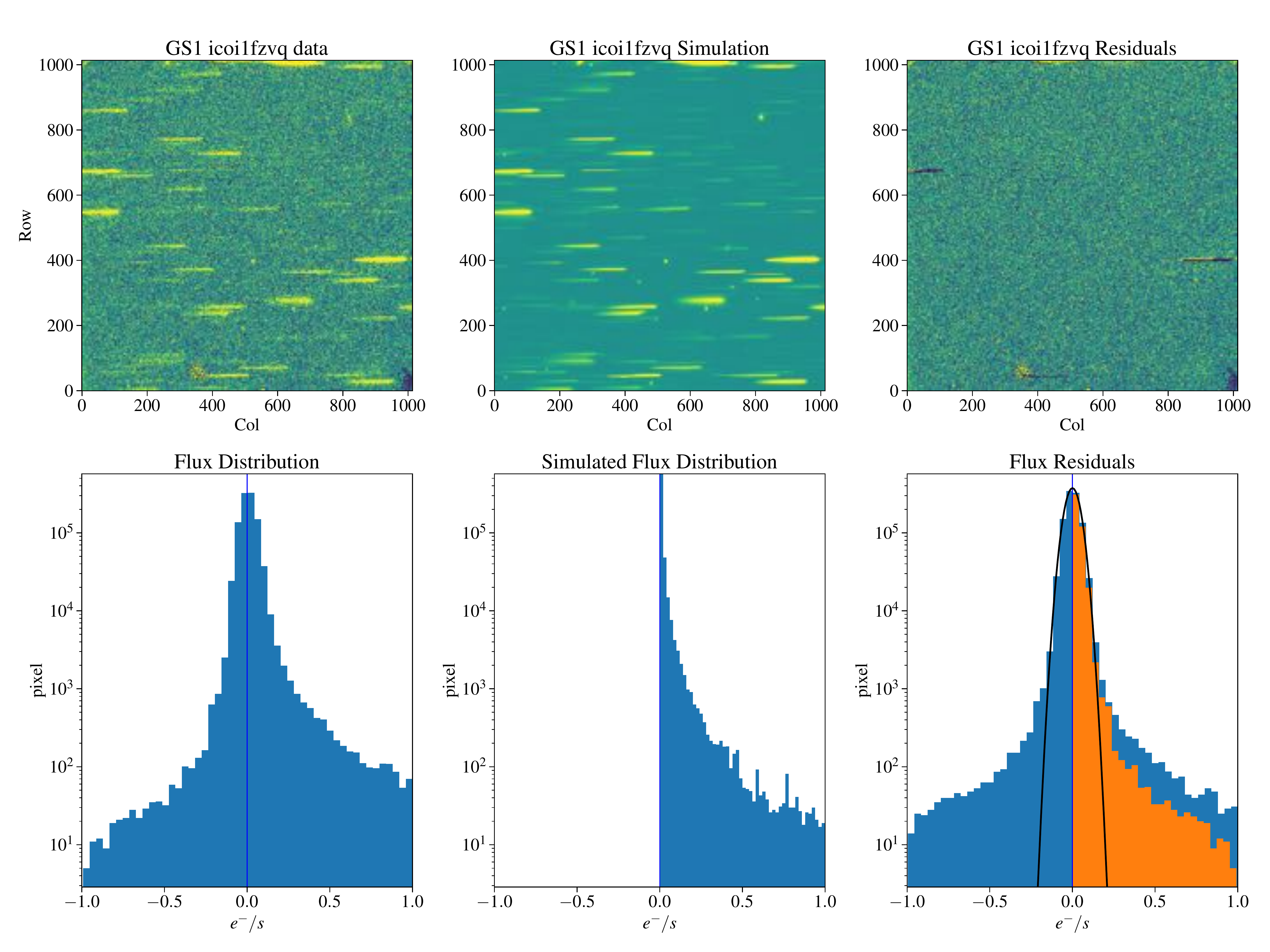}
\caption{Same as Figure \ref{FIGS_simul_GN1} but for the GS1 field. \label{FIGS_simul_GS1}}
\end{figure*}

\begin{figure*}
\center
\includegraphics[width=7.in]{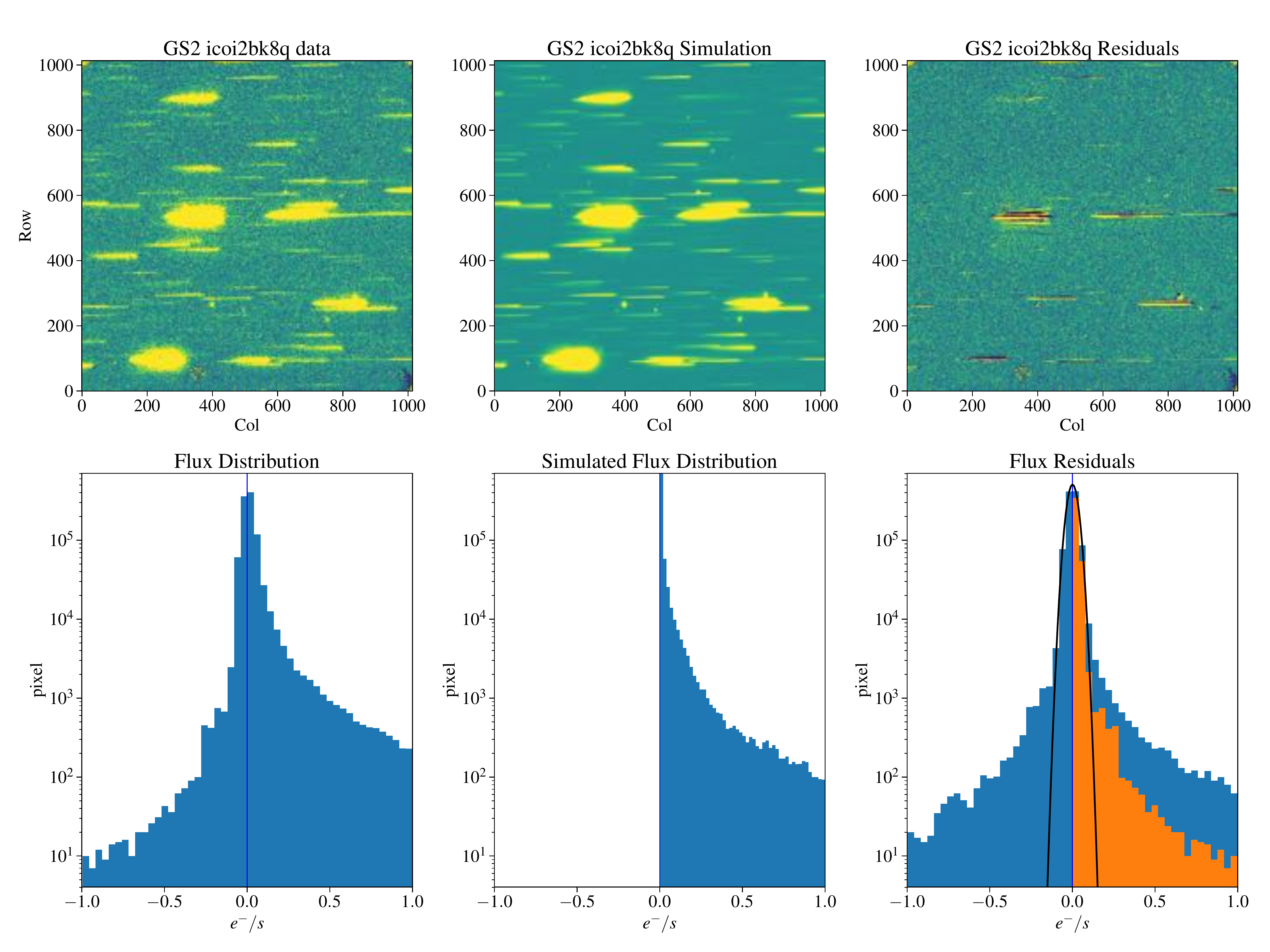}
\caption{Same as Figure \ref{FIGS_simul_GN1} but for the GS2 field. \label{FIGS_simul_GS2}}
\end{figure*}

When analyzing slitess data, one must necessarily have some knowledge of where each source is expected to be in the field, as well as have a good understanding and calibration of the WFC3 G102 grism \citep{Pirzkal2016}. Since spectra of different objects overlap in our observations, we must be able to estimate where light for every single source in the field will be dispersed to for all five different spectral orders of the G102 grism. While we are only interested in extracting data from the first dispersed order, a complete tally of the contamination from other objects and spectral orders is required. We used the publicly available WFC3 G102 grism calibration file  \citep{Pirzkal2016,Pirzkal2017a} with new custom software to simulate every single FIGS grism observations. These simulations were based on the broad band continuum SED of objects in the field. While bright emission lines in higher spectral orders are not simulated at this stage, the use of data taken at multiple position angles is sufficient to exclude such lines or other artifacts are bonafide emission lines in an observed first order spectra.
The software dispersed every object-pixel in the master mosaics onto the reference frame of the FIGS observations, using the multiple broad-band mosaics to assign a spectral energy distribution to each pixel. For each pixel, the broad band fluxes are interpolated to form a smooth SED. This process is similar to how the {\ tt aXe} \citep{pirzkal2001,pasquali06, Kummel2009} software package models observations.  However, it allowed us to compute the dispersion solution only once, and store this information for later use when extracting data or estimating the contamination level in a spectrum. The SBE approach produces large tables containing for each grism pixel in every datase a list of what object-pixel contributed what amount of flux and at what wavelength. These large data cubes can be used for a variety of tasks, such as generating simulated dispersed images for individual objects or all objects. These data cubes can also be used to determine which pixel in the real observations need to be extracted and co-added to produce ''rectified'' 2D images, where the x-axis is now a linear function of wavelength. This approach is also well suited to generate forward modeling models of star forming regions, as will be presented in a forthcoming paper \citet{Pirzkal2017c}.
One additional advantage of the SBE approach is that we are able to compare simulations to observations early in the analysis process. When the astrometry and input catalogs are sufficiently accurate, the simulations and the observations should be very similar. Thus, it is a confirmation of the accuracy of any future individual object contamination estimates, as well as any future extracted ''rectified'' 2D image of individual objects in the field.
In Figures \ref{FIGS_simul_GN1},\ref{FIGS_simul_GN2},\ref{FIGS_simul_GS1}, and \ref{FIGS_simul_GS2}, we compare a single FIGS observation to its FIGS simulation. The lack of significant shift in the x or y direction (except for the brighter stars with detectable proper motion) indicates a good astrometric solution. The quality of the continuum level subtraction is an indication that our estimate of the SED of each pixel was also accurate. Regions with poor subtractions are caused by saturation of bright objects, persistence effects, and proper motions of some of the bright stars in the fields.

\subsubsection{Persistence Flagging}
Persistence \citep{knox2013} affects many WFC3 IR exposures. In the case of FIGS data, it can be caused by previous FIGS exposures themselves, or by other HST programs executed before a given FIGS exposure. We used the HST archive to identify every exposure obtained within 36 hours of a FIGS exposure, and flagged any pixels that were saturated within that period. Typically, at most, a few hundred pixels were affected in a particular image. Flagging these pixels was essential since persistence could mimic faint emission lines.

\subsubsection{Background Subtraction}\label{back_sub}
\begin{figure*}
\center
\includegraphics[width=7.in]{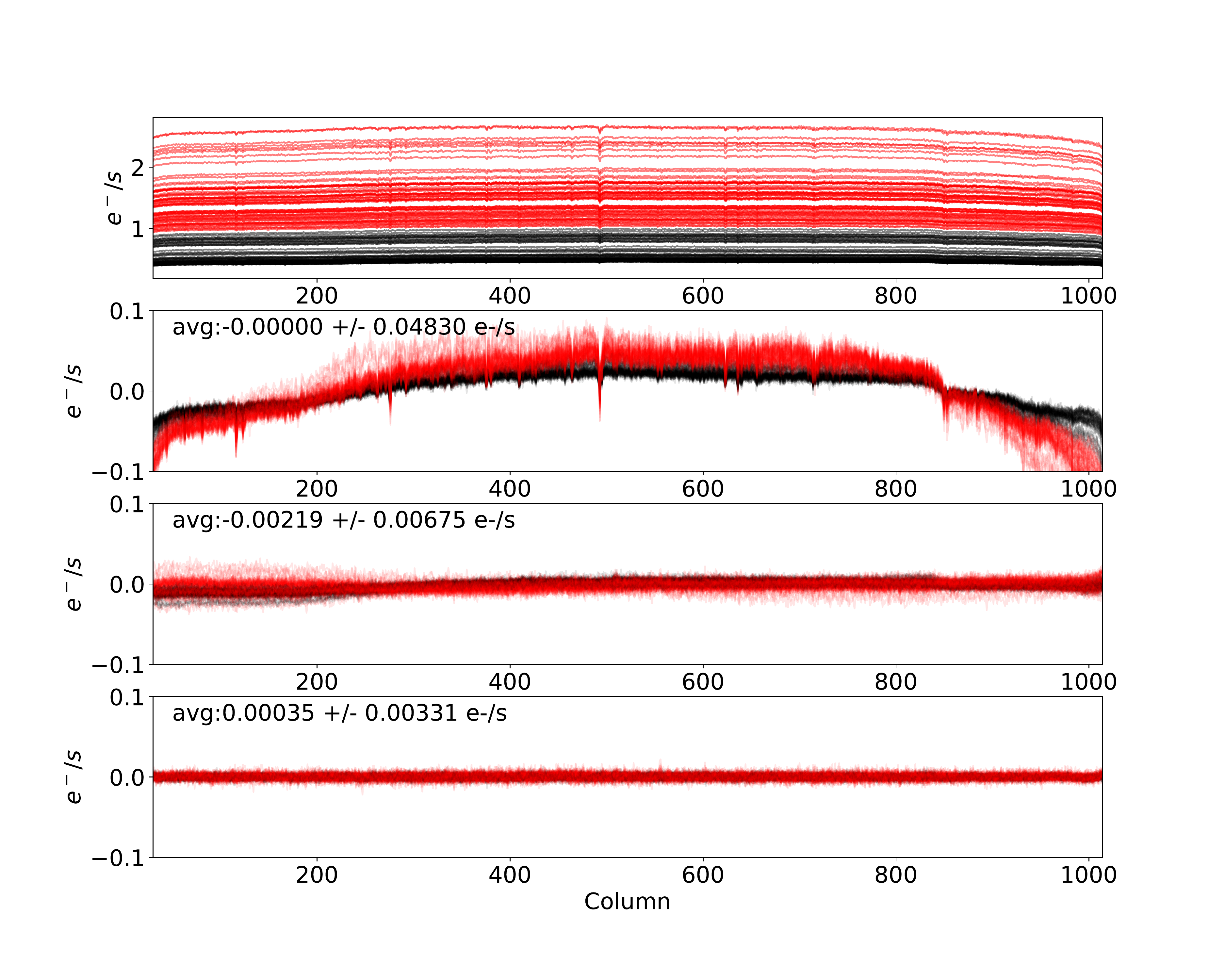}
\vspace*{-1.0cm}
\caption{This Figure shows the median column values, as a function of column, of all the observations taken for the FIGS GS1 field. The top panel shows the original background levels of the observations. Observations with a large amount of HeI light are plotted in red. The second panel shows the mean subtracted background residuals if a constant dispersed background is used, and shows how observations with a large amount of HeI have a different background structure. The third panel show the result of implementing the two components, and varying HeI level background subtraction described in Section \ref{back_sub}. Finally, the fourth panel shows the residuals when an additional smooth component is fitted and subtracted. As we show here, the FIGS background subtraction brings the background level of our grism observations to within $0.003 e^-/s$\ of zero.\label{GS1_back}}
\end{figure*}

The dispersed background light must be subtracted from slitless observations, so that any under or over subtraction of the background sets a limit to our ability to detect faint continuum or emission lines. The dispersed background not only has spatial structure imposed by the multiple overlapping spectra orders, but can also vary substantially during the course of an observation. This is because the bright Earth limb contributes to the overall background light in the form of HeI emission from the Earth's upper atmosphere. At the beginning and end of an observation, HST is more likely to be pointing close to the Earth's limb, and to be affected by HeI emission.
Unfortunately, the default HST WFC3 data calibration pipeline (CALWF3) cannot handle images with varying background properly when up-the-ramp fitting (UTRF) is used \citep{Robberto2007}. CALWF3 uses UTRF of multiple non-destructive reads to produce images with a lower effective read-noise ($12 e^-$\ versus $20 e^-$) and which is free of cosmic ray impacts. 
To deal with this significant issue, we assumed that the dispersed grism background is a linear combination of two separate contributions: the regular and constant dispersed Zodiacal light, and a time varying dispersed HeI background. These two components were previously modeled and are described in \citet{Brammer2015}. 
We then estimated the amount of HeI light separately in each of the multiple reads taken during the exposure. This HeI estimate was then subtracted from each of the reads, thereby removing the varying component from the observation, UTRF was then used to produce a set of intermediate images. The Zodiacal light was then subtracted from these images to produce a final set of images. This method is described in detail in  \citep{Pirzkal2017b}. While observations typically have background levels of $\approx 0.5 - 1.4 e^-/s$\ with spatial variations of $\approx 0.05 e^-/s$, this method yields images with a background residuals of  $0.002 \pm 0.005 e^-/s$. This was further improved by computing the median of an observation in the row direction, and then smoothing and subtracting it from the data.  The first step in this further refinement was to mask pixels affected by a level of persistence that is 0.6 times higher than the error estimate, and then to mask pixels affected by dispersed spectra with a count rate larger than $0.01 e^-/s$ (as determined in our simulated images discussed in Section \ref{simulations}). This final step further reduced the background residuals to a level of $\pm 0.003 e^-/s$. Figure \ref{GS1_back} shows the result of this process for the FIGS GS1 observations.

\subsection{Extraction}
\subsubsection{2D extraction}\label{2Dextract}
\begin{figure*}
\center
\vspace{-1.0in}
\includegraphics[width=6.in]{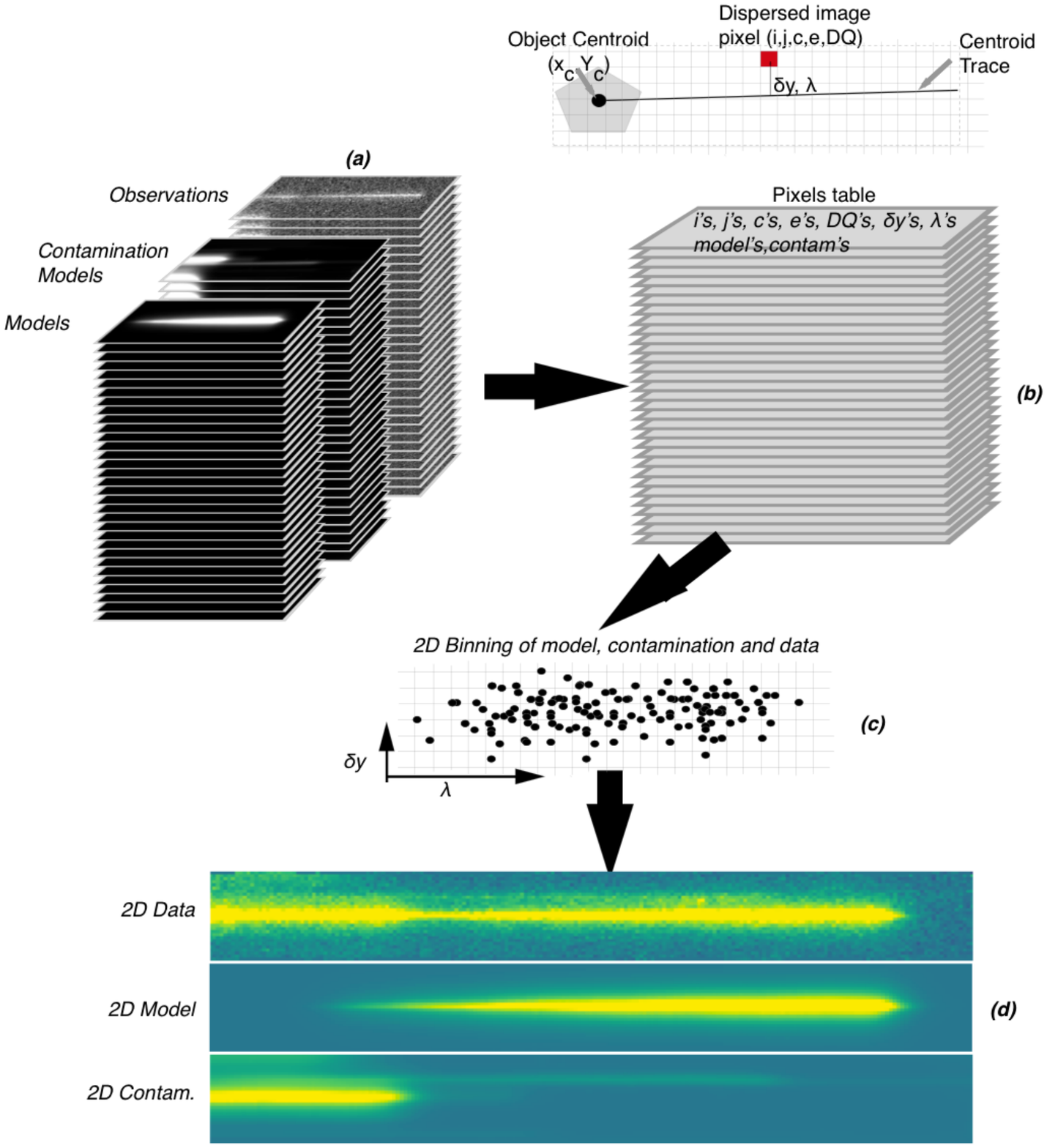}
\vspace*{-1.0cm}
\caption{This Figure illustrates the extraction process described in Section \ref{2Dextract}. Stacks of observed data, and models (Panel a) combined with the known field-dependent G102 grism dispersion are used to generate the pixel tables (Panel b). The pixel table contains for each pixel: the detector coordinates (i,j), flux (c), flux error (e), data quality flag (DQ), distance of center of pixel to the trace ($\delta y$), and assigned wavelength ($\lambda$),  The latter can be used to map the observations, model estimates as well as contamination, in a wavelength versus cross-dispersion distance ($\lambda$\ vs.  ${\rm \delta y}$) space (Panel c). Whence mapped into this two dimensional space, the flux (c) can be binned two-dimensionally in $lambda$\ and ${\rm \delta y}$. Flux values in each bins are combined using a $3\sigma$\ clipping weighted mean, using the original error estimates (e) of the original G102 fluxes (c). The standard deviation of this mean is computed and serves as a robust and empiracal estimate of the error in each bin. This binning process results in combined 2D-images of the flux ,error , model, and contamination estimates that are wavelength calibrated (d) and rectified in the cross-dispersion direction. \label{FIGS2Ddiag}}
\end{figure*}

Each object present in the SeXtractor segmentation file described in Section \ref{catalogs} was extracted. First, for each available {\it FLT} file, the dispersed trace corresponding to the centroid coordinates listed in the FIGS catalog was computed. Then, pixels in the {\it FLT} data (avoiding those flagged as bad in the {\it FLT} Data Quality extension)  were assigned a wavelength ($\lambda$) and the cross dispersion distance ($\delta$y)  between the dispersed centroid trace and the center of that pixel was computed. We also used the simulated data cubes discussed in Section \ref{simulations} to determine the amount of flux in that pixel contributed by the spectra of other sources in the field. These three quantities supplement the already known observed count rate, error estimate, and data quality flag of that pixel. This is very similar to what the {\tt aXe} extraction software does \citet{Kummel2009}. However, our approach differs in that, unlike {\tt aXe}, this information was not used to produce individual {\it FLT} extraction of the spectrum of each object. Instead, the information obtained using {\em all} of the {\it FLT} data obtained at the same PA on the sky (i.e. 16 or 32 of them) were combined. 
A two dimensional, wavelength rectified image was then generated by binning those data in  $\lambda$\ and $\delta$y space. 
The bins were chosen to be 25\AA\ in the wavelength direction and one WFC3 pixel
in the $\delta$y direction ($0.129$\arcsec), as these are close to the native properties of the instrument. 
Since we had data from multiple, dithered {\it FLT} files to bin, the number of available sample in each bin was sufficient to both derive a robust ($3\sigma$\ clipping mean) estimate of the count rate in each bin, as well as to derive an accurate standard deviation of the mean for each of the two dimensional bins. The identical process was followed to generate 2D-rectified images of the simulated spectrum of each source, as well as of the simulated contamination estimate for each source. This process is illustrated in Figure \ref{FIGS2Ddiag}.

\subsubsection{1D extraction}
\begin{figure*}
\center
\includegraphics[width=6.in]{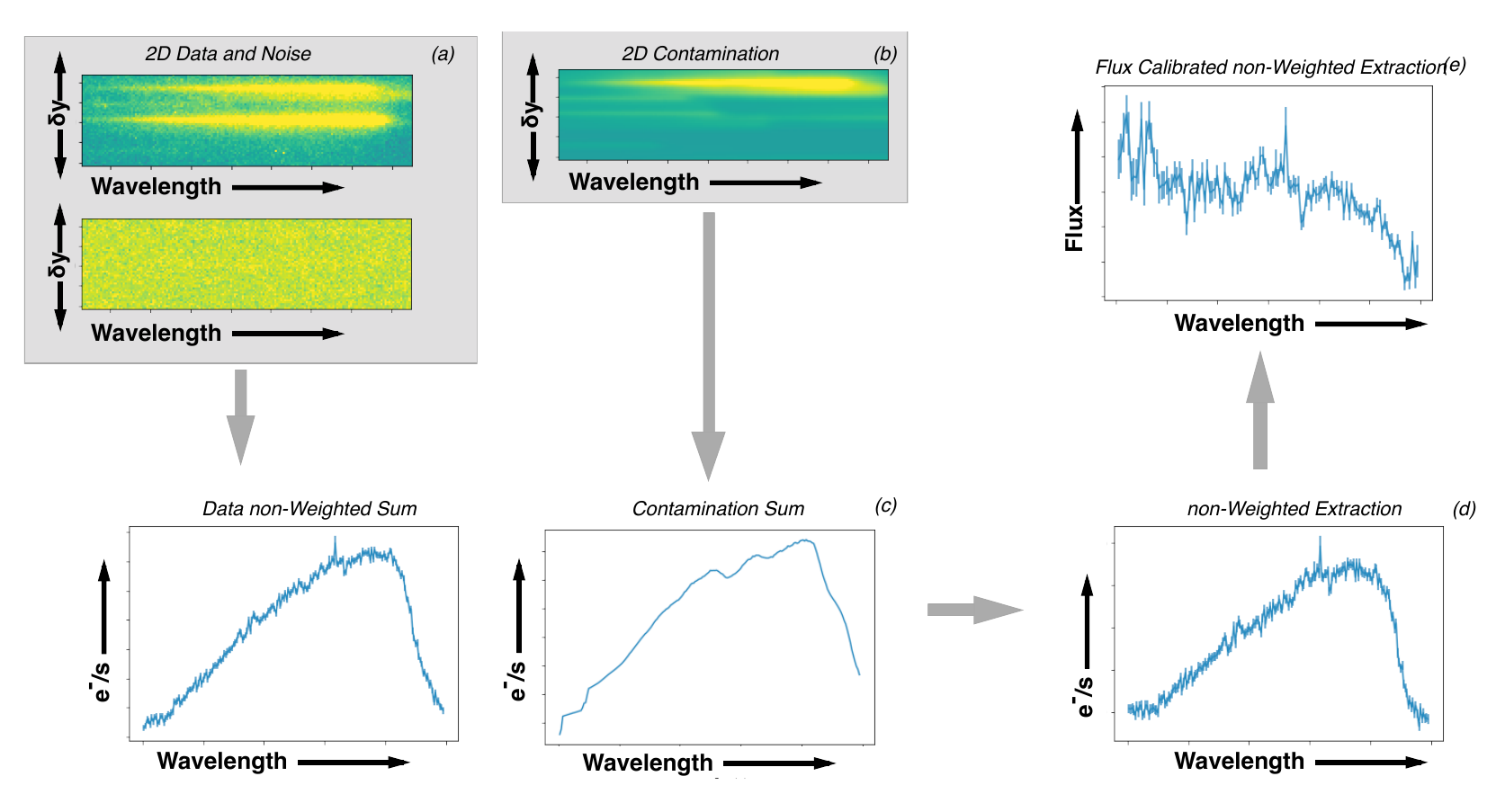}
\caption{FIGS Non-weighted Extraction process: The data and error 2D data: (a) are summed and errors are propagated to produce a 1D spectrum of the source. The same process is used with the 2D contamination estimate (b) for this object to produce a 1D spectral estimate of the contamination (c). The 1D contamination estimate is subtracted from the 1D source spectrum to produce a contamination-free spectrum (d). The final flux-calibrated 1D spectrum is then generated by applying the grism sensitivity (e). \label{FIGSsum1D}}
\end{figure*}

\begin{figure*}
\center
\includegraphics[width=6.in]{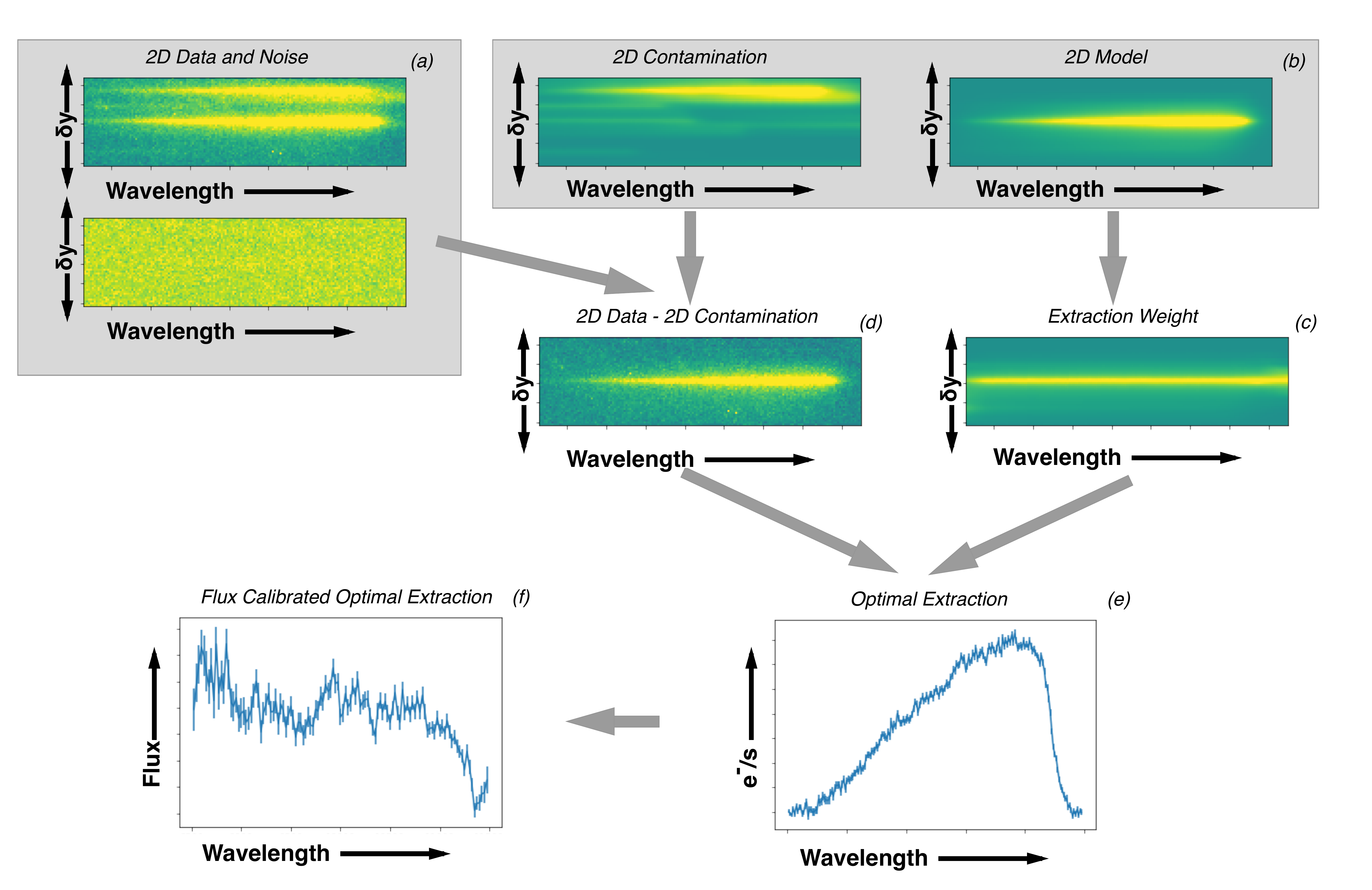}
\caption{FIGS Optimal Extraction process: The 2D contamination (b) is first subtracted from the 2D data (a) to produce a contamination free 2D estimate (d). The 2D model of the dispersed spectrum of the source (b) is used to create a normalized extraction weight image (c). The optimal extraction algorithm from \citet{Horne1986} combined data, error, and extraction weight to produce a 1D estimate of the spectrum of the source (e).  The final flux calibrated 1D spectrum is then generated by applying the grism sensitivity (f).  \label{FIGSopt1D}}
\end{figure*}

One dimensional extractions were created from the 2D-extractions using two methods: Non-weighted extraction and Optimal extraction. These methods are shown graphically in Figures \ref{FIGSsum1D} and \ref{FIGSopt1D}.
In the first instance, rows of the 2D-rectified spectra (Figure  \ref{FIGSsum1D}a) are simply co-added. The SeXtractor segmentation footprint determines which rows should be included in this summing operation. The 2D model of the spectral contamination (Figure  \ref{FIGSsum1D}b) are also  co-added to create a 1D spectral estimate of the contamination. The latter is subtracted from the 1D spectrum of the source (Figure  \ref{FIGSsum1D}c) and a final 1D spectrum is generated by applying the known grism sensitivity (Figure  \ref{FIGSsum1D}e).

The FIGS optimal extraction follows a non-iterative version of the algorithm described in \citet{Horne1986}: We used the simulated version of the 2D dispersed spectrum of the source to determine the expected profile of the spectrum as a function of wavelength (Figure  \ref{FIGSopt1D}b). This profile was normalized to unity in the cross dispersion direction and used as the extraction weight (Figure  \ref{FIGSopt1D}c). This extraction weight was then used in combination with the 2D contamination subtracted 2D data (Figure  \ref{FIGSopt1D}d), to produce an optimally extracted 1D spectrum (Figure  \ref{FIGSopt1D}f).
The optimal extraction has the advantage of producing higher S/N spectra with improved flux calibration, but only when the extraction weights (derived from the imaging data) are accurate. This is not always the case, as in the example of stars with proper motion. In such cases, the extraction weights are misaligned and cause spectral artifacts in the extracted data. Checking the consistency between the co-added and optimally extracted spectra is {\it always} recommended.

\section{FIGS Spectra}\label{sec:spectra}

\subsection{Depth of Complete Survey}
\begin{figure}
\center
\includegraphics[width=3.5in]{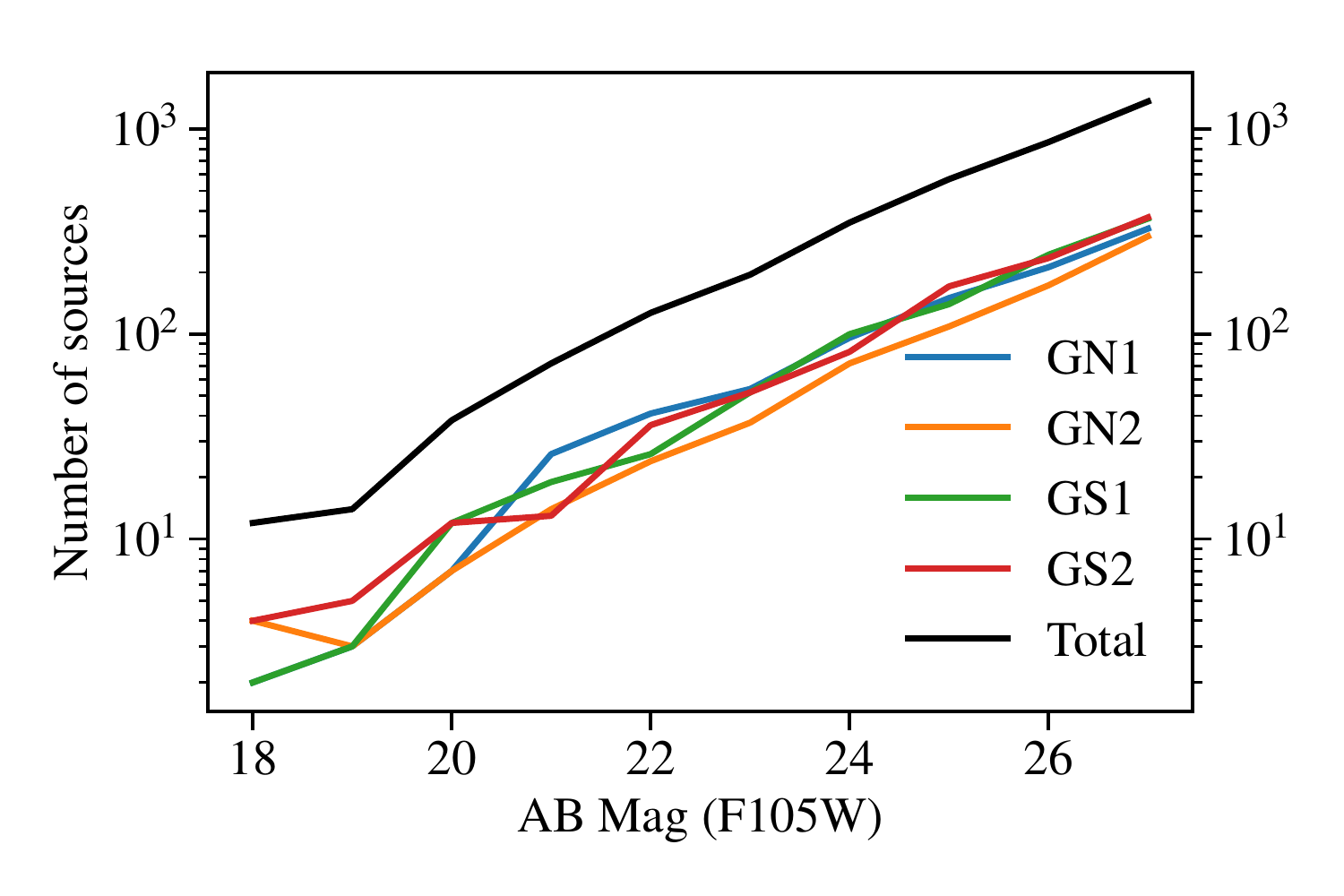}
\caption{
Number of source spectra as a function of source $m_{F105W}$\ magnitude for each FIGS field and for the total FIGS sample (black). \label{FIGSSN}}
\end{figure}

The number of FIGS spectra, up to five per source, as a function of broadband $m_{F105W}$\ magnitude is shown in the right panel of Figure \ref{FIGSSN}.
The total number of sources extracted in each of the FIGS fields were determined by the depth of the available imaging data and the size of the catalog for each field, as well as the specific position angles.  The total number of extracted sources for the GN1, GN2, GS1, and GS2 fields are 1913, 1003, 3106, and 2623, respectively. The increased number of extracted spectra in the GS1 and GS2 is due to the deeper images and larger object catalogs available for these fields. The number of {\it sources} brighter than $m_{F105W}=26.5$\ mag that were extracted are 595, 453, 603, and 619 for GN1, GN2, GS1, and GS2, respectively, The total number of {\it extracted spectra} for all four fields brighter than $m_{F105W}=26.5$\ mag is 2270.

\begin{figure}
\center
\includegraphics[width=3.5in]{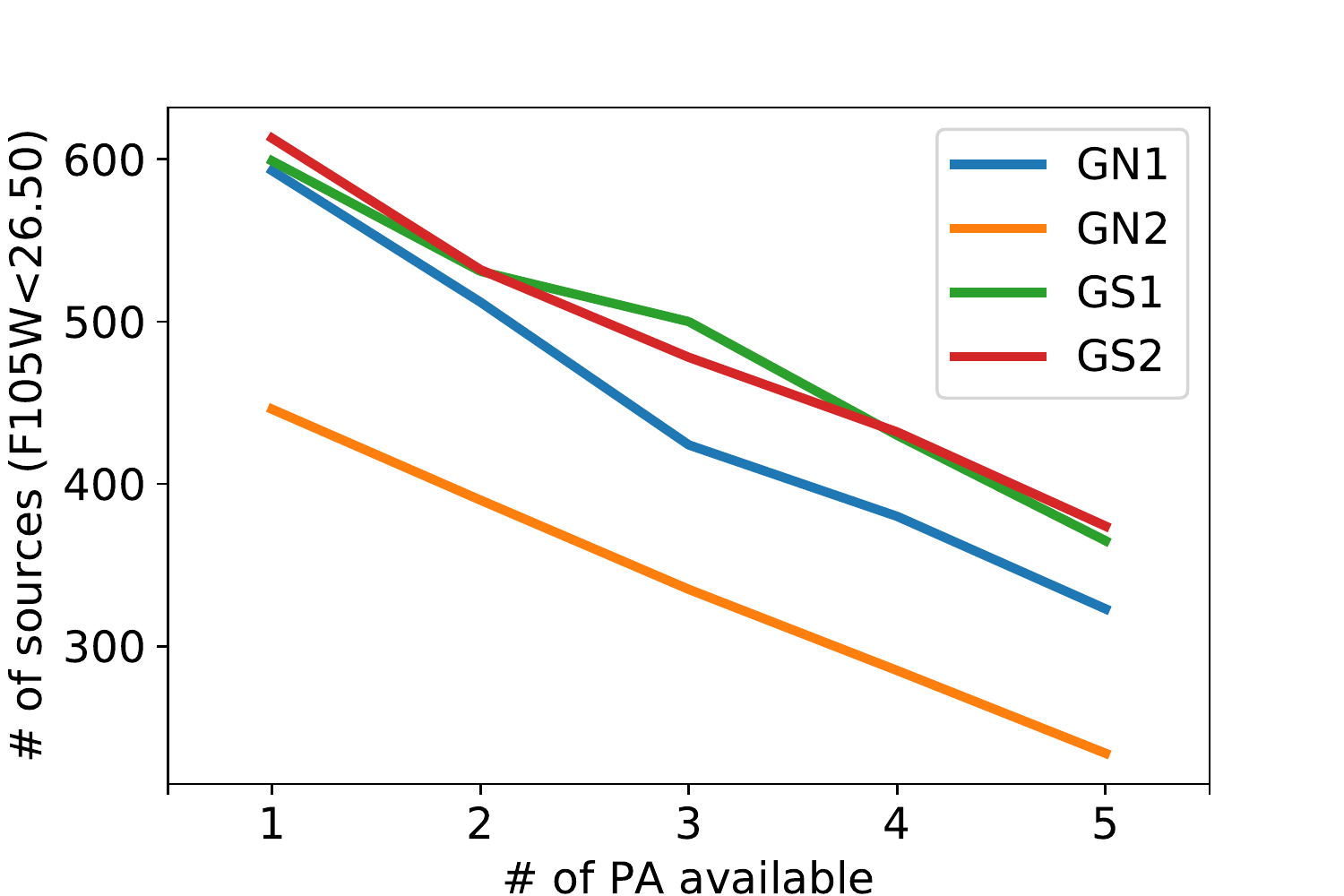}
\caption{Number of objects with  $m_{F105W}<26.5$\ magnitude , observed in 1, 2, 3, 4 or 5 position angles. Approximately 55\% of all of the sources were observed to the full depth of 40 orbits. \label{FIGSdepth}}
\end{figure}

Not every source was observed to the full depth corresponding to 5 position angles, and Figure \ref{FIGSdepth} shows the number of sources brighter than $m_{F105W}=26.5$\ mag which where observed in at least $n$\ position angles. Approximately 57\% of the 2270 sources brighter than $m_{F105W}=26.5$\ mag were observed to the full depth of the survey, or 323, 234, 365, and 374 sources in GN1, GN2, GS1, and GS2, respectively.


\subsection{Combined Spectra} \label{sec:combined}
Each source in the FIGS survey was observed up to five times, producing five distinct spectra. Due to the properties of slitless spectroscopy, spectra obtained for even moderately extended object are subject to a different amount of smoothing. Thus, the resolution is set by the image of the object itself.

\begin{figure}
\includegraphics[width=3.5in]{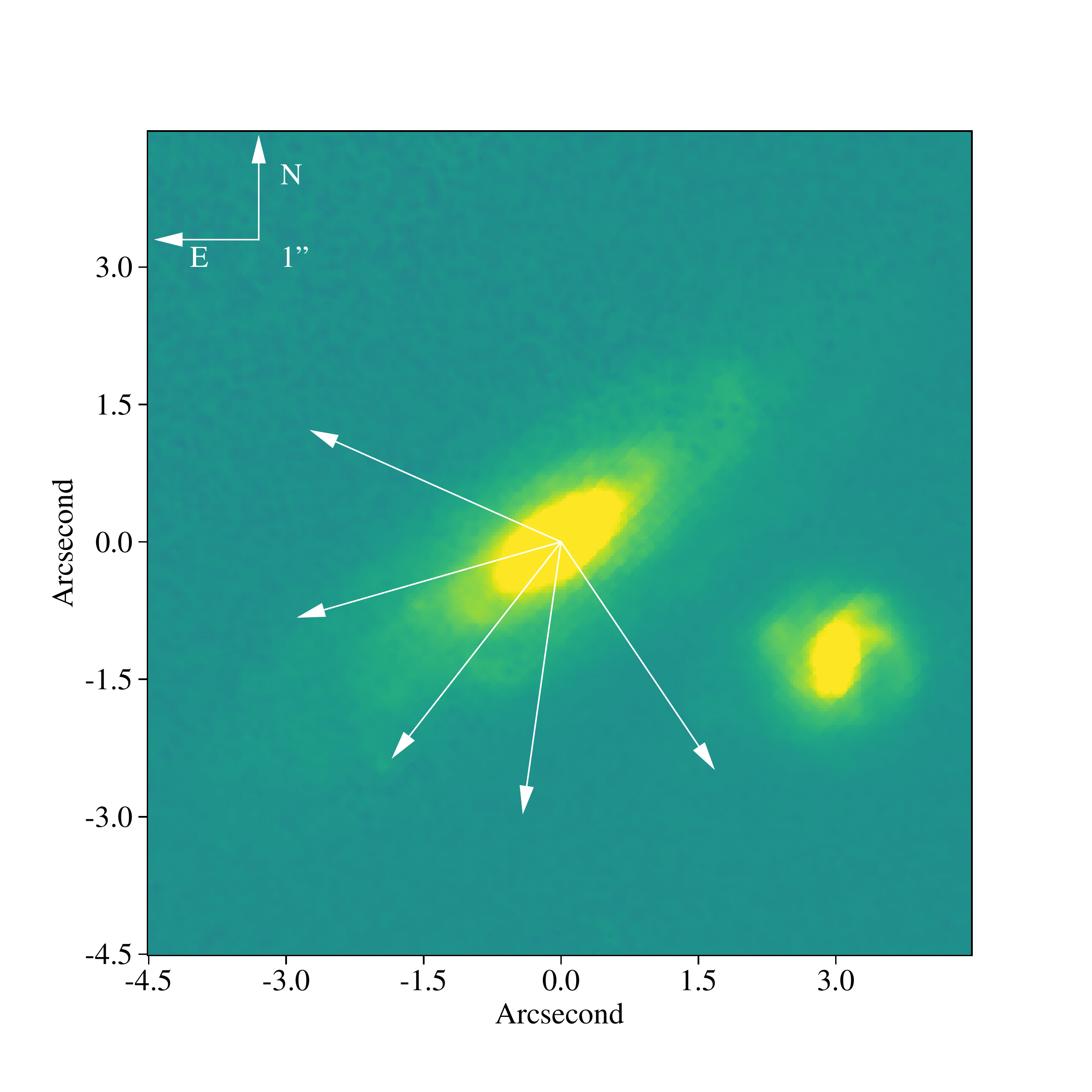}
\caption{An extended, asymmetrical galaxy in the GN1 field, and the 5 directions along which spectra were obtained.\label{FIGS_1496_dir}}
\end{figure}

\begin{figure*}
\center
\hbox{
\hspace{-0.3in}
\includegraphics[width=3.5in]{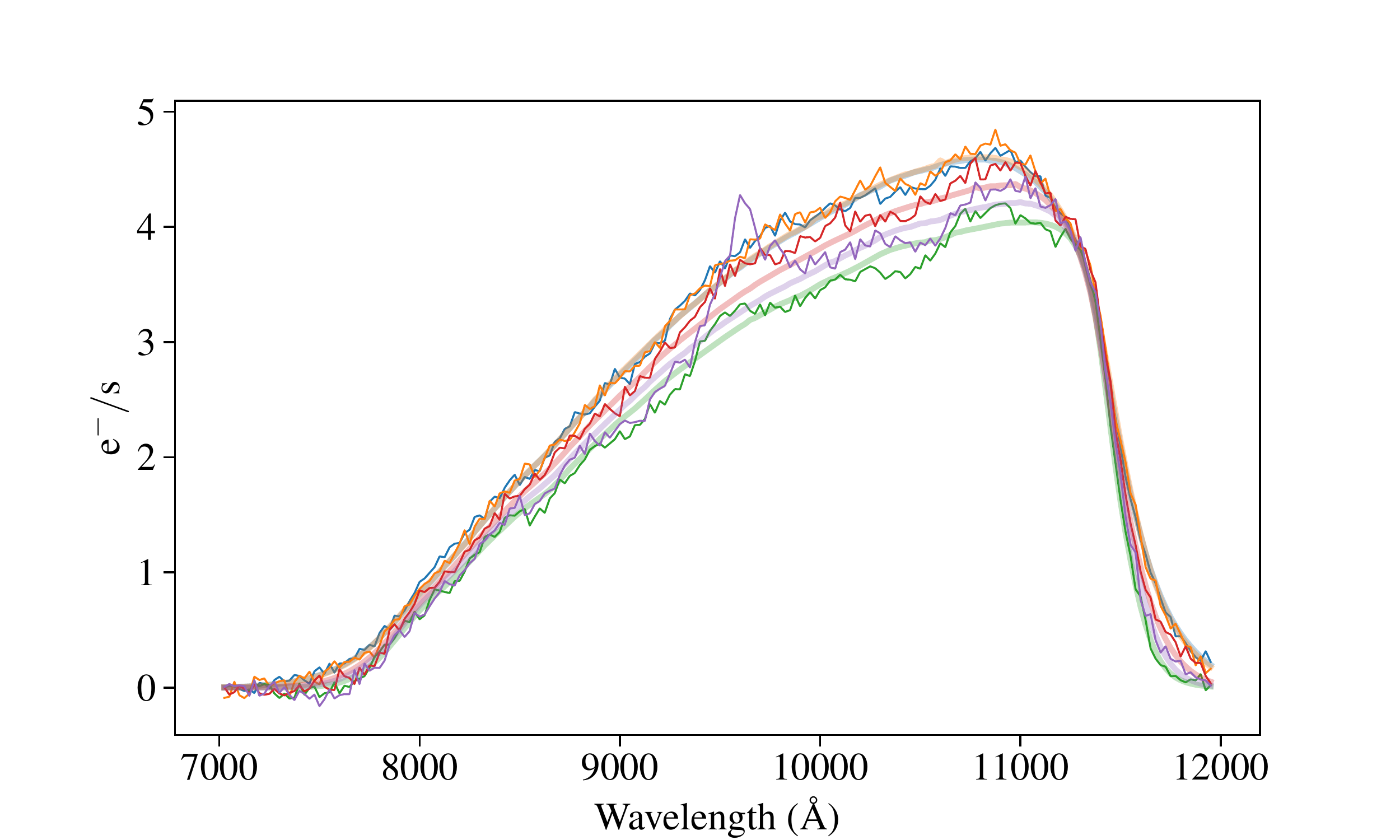}
\hspace{-0.3in}
\includegraphics[width=3.5in]{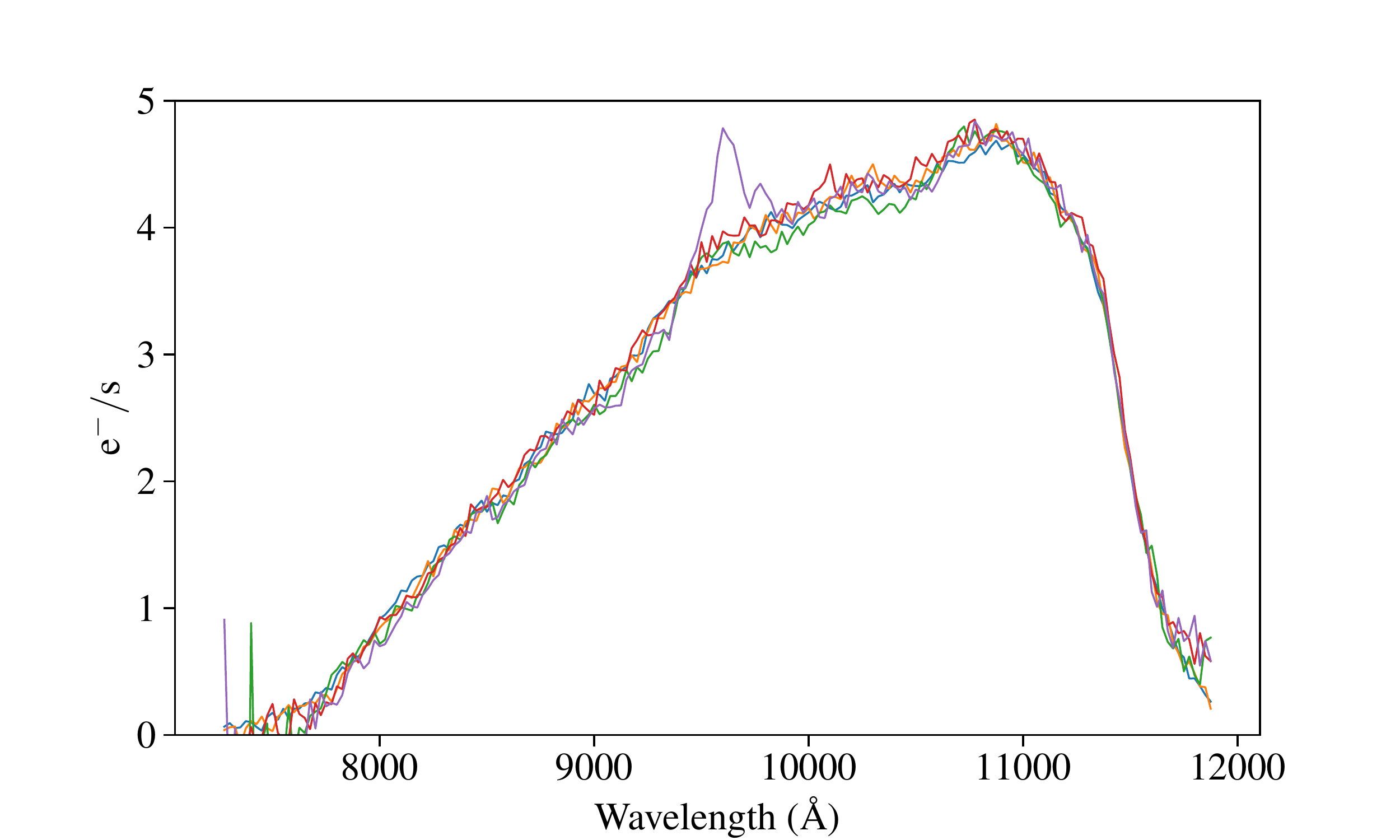}
}
\caption{Left: Non-flux calibrated 1D spectra of the large galaxy shown in Figure \ref{FIGS_1496_dir} as well as the FIGS model of the same galaxy. The large asymmetry of the galaxy results in significant different amount of smoothing, which results in wider as well as overall lower amplitude spectra. Right: Same spectra as shown in the Left Panel, but an object specific correction based on the FIGS extraction of the simulation of this object was applied.\label{FIGS_GN1_1496}}
\end{figure*}

\begin{figure*}
\center
\hbox{
\hspace{-0.3in}
\includegraphics[width=3.5in]{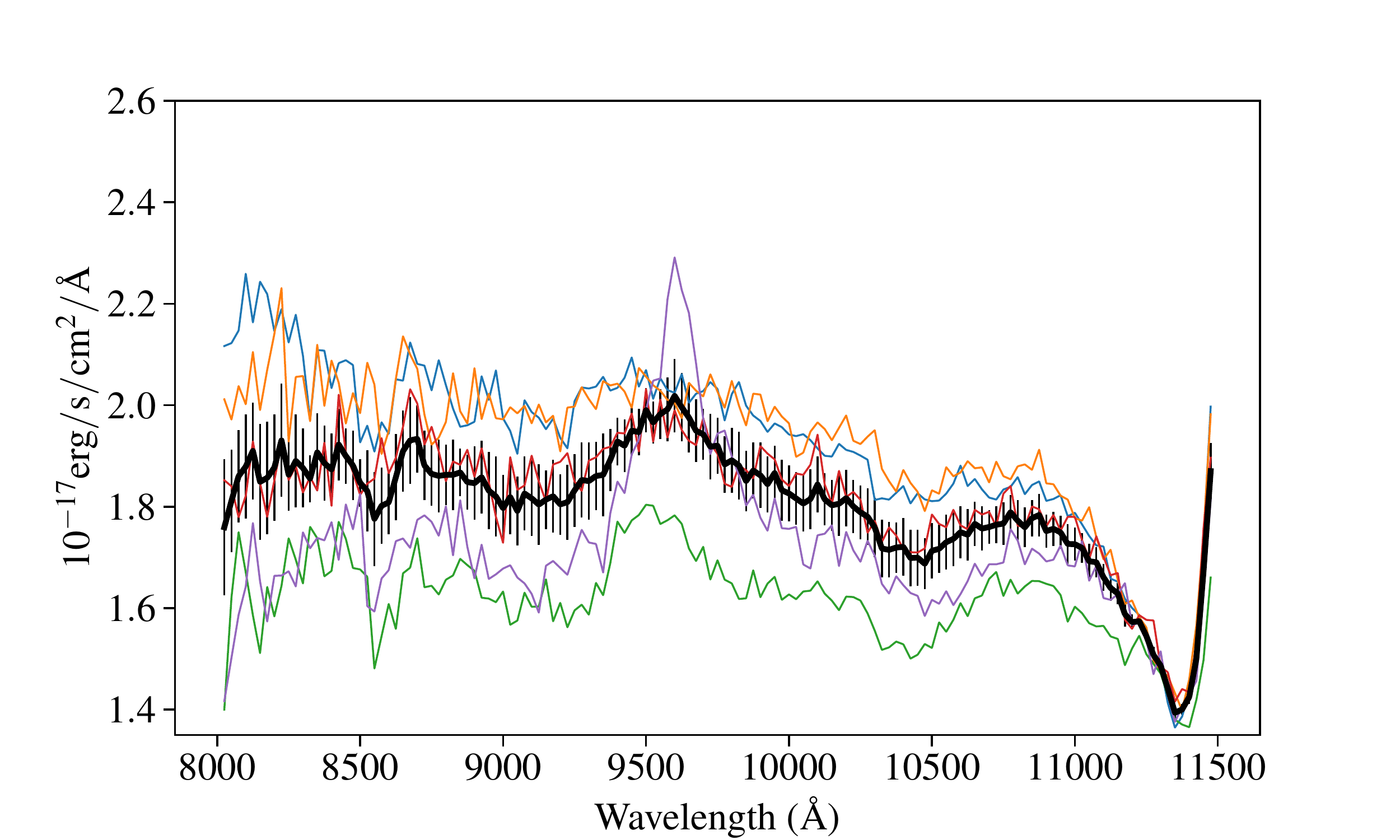}
\hspace{-0.3in}
\includegraphics[width=3.5in]{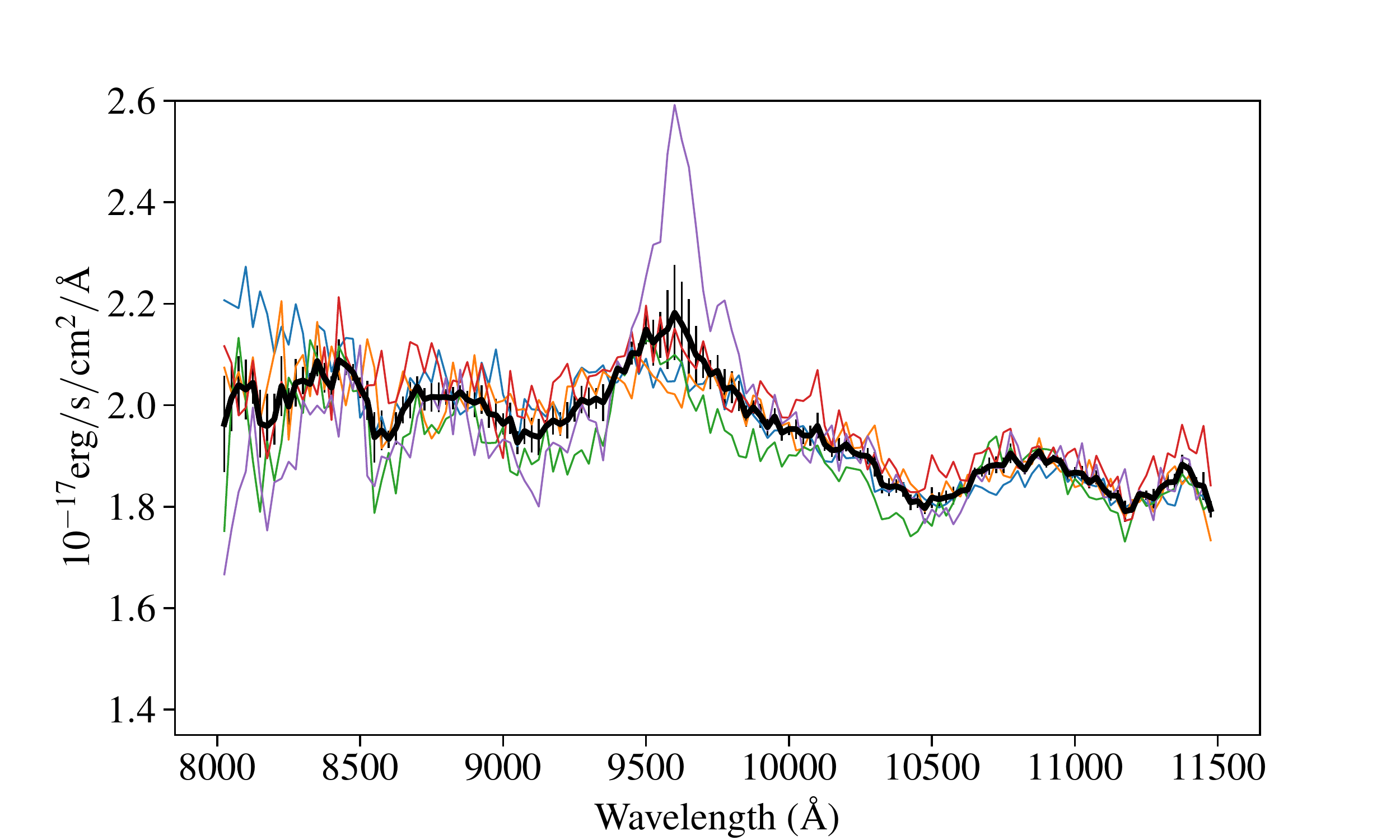}
}
\caption{Left: Same as the Left panel of Figure \ref{FIGS_GN1_1496}, but after applying the default G102 spectral sensitivity function. We also show the result of combining these spectra (black line with error bars). Right: Flux calibrated spectra of this source after applying an object-specific G102 spectral sensitivity function. The resulting combined spectrum (shown in black) is significantly less noisy, and does not suffer from the edge effects of the combined spectrum shown in the Left panel.The broad spectral feature at $9500\AA$, which could be the result of some extended H$\alpha$\ emission, is readily visible in the final combined spectrum. \label{FIGS_GN1_1496_flx}}
\end{figure*}

The observed spectra are the convolution of the light profile of the object with its spectrum, and large differences in this light profile between different PA (for example, in the cases of elliptical or irregular galaxies) will result in spectra that disagree strongly near the edge of the bandpass of the grism. They will also have continuum fluxes that are in disagreement, as the spectrum is smoothed by different amounts. 
This effect is illustrated in Figures \ref{FIGS_1496_dir}, \ref{FIGS_GN1_1496} and  \ref{FIGS_GN1_1496_flx}, where we show the example of a large  galaxy.
We derived an object-specific spectral response for each source by dividing the extracted 1D data by the extracted 1D simulated data, and by the spectral energy distribution used to generate the FIGS simulations, which were generated from the available FIGS broad band photometry. The result is a normalized spectrum , which can be scaled back to the observed F105W photometry. These steps insure that the 1D spectra of extended sources are accurately flux-calibrated and  avoid the issue of having a point-source sensitivity function applied to an extended object.

The FIGS spectra were flux calibrated using object specific sensitivity functions and then combined. For each wavelength bin, the inverse variance of the single-PA spectra were used as weights to compute the weighted mean and standard deviation of the weighted mean. An iterative $3\sigma$\ rejection was used to remove outlier single PA spectral bins. 


\subsection{Net Significance}

The information content of a spectrum can be better described by the Net Significance ($\mathcal{N}$), which was introduced in \citet{pirzkal04}. $\mathcal{N}$\ is the maximum cumulative S/N of a spectrum. We compute it as follow, for each spectrum:
\begin{itemize}

\item Divide the $\approx$\ 140 flux values $F_i$\ by their respective error estimates $e_i$\ to produce 140 S/N estimates $SN_i$.
\item Re-order the original flux and error arrays according to the descending order of the S/N estimates to produce $F'_i$\ and $e'_i$.
\item Compute $\mathcal{N} = \sum_{i=0}^{n} F'_i /  \sqrt {\sum_{i=0}^{n} (e'_i)^2}$\ for increasing values of $n$,  until the maximum value of $\mathcal{N}$\ is reached. 
\end{itemize}

\begin{figure*}
\center
\includegraphics[width=3.5in]{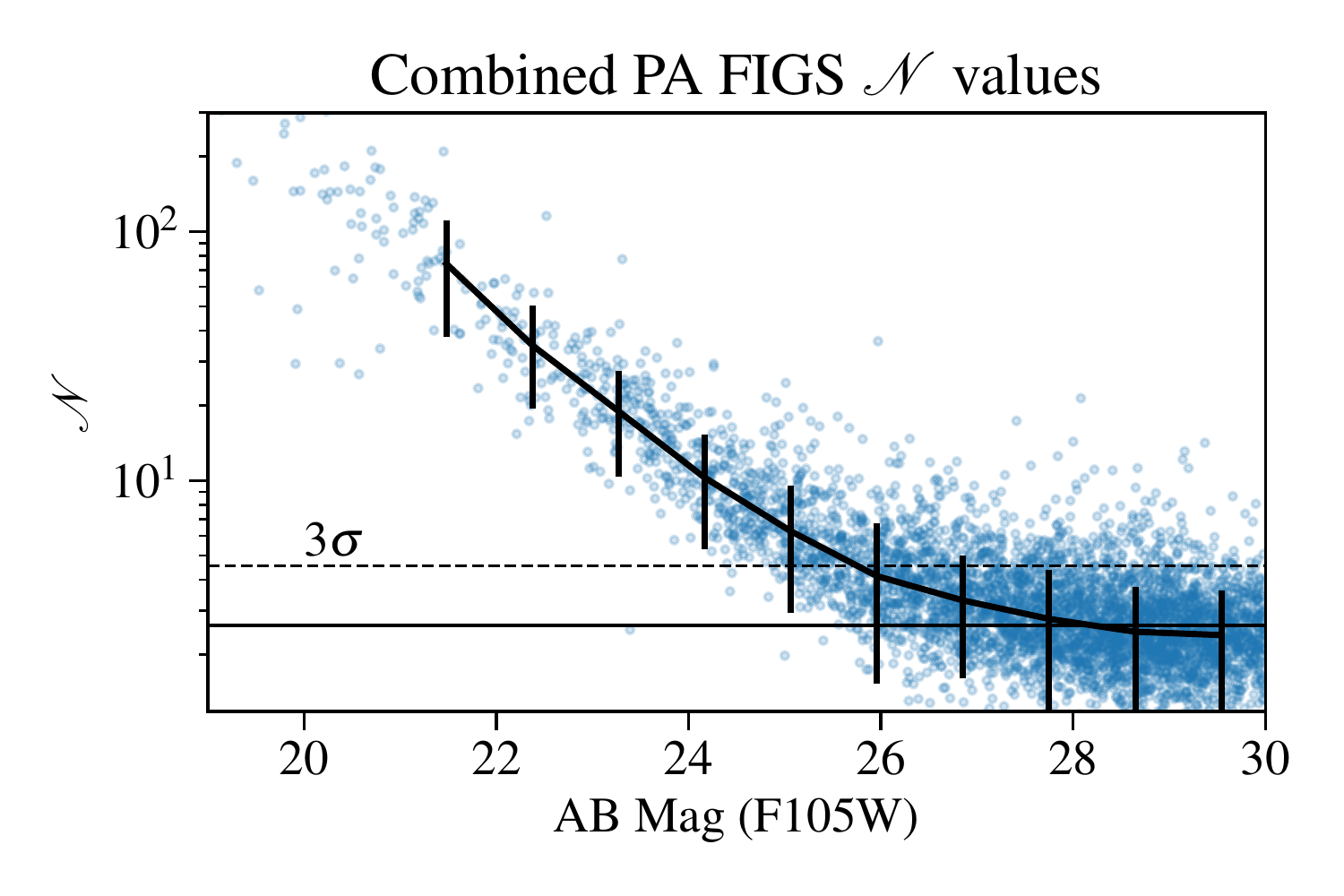}

\caption{The distribution of $\mathcal{N}$ values for combined PA-depth FIGS spectra (blue).The mean value of $\mathcal{N}$\ is shown in black with error bars and approaches the value of $\mathcal{N}\approx2.6$ (lower horizontal black line) for source that are several magnitudes fainter then our detection limit. The $3\sigma$\ continuum level at a limit $\mathcal{N}\approx4.5$\ is also shown (dashed line).  \label{FIGSNfaint}}
\end{figure*}

Figure \ref{FIGSNfaint} shows a plot of the computed $\mathcal{N}$\ for the combined five PA depth FIGS spectra. 
Sources fainter than $m_{F105W}=28$\ are below our G102 grism sensitivity and Figure  \ref{FIGSNfaint} shows that $\mathcal{N}$\ approaches its limiting value of 2.6. A simulated spectrum with a continuum level of $3\sigma$\ per bin, is expected to have $\mathcal{N}\approx4.5$.
There are several factors that can of course affect a value of $\mathcal{N}$ and unaccounted contamination or detector artifact can result in an artificially high value of $\mathcal{N}$. Similarly, small error in the background subtraction can raise or lower the level of any continuum signal, affecting the value of  $\mathcal{N}$. As a whole, the $\mathcal{N}$\ distribution shown in Figure \ref{FIGSNfaint} indicates that, on average, the $3\sigma$\ detection limit of the FIGS survey corresponds approximately to a $m_{F105W}=26$\ continuum source.

\subsection{Examples} \label{sec:examples}
\begin{figure*}
\center

\includegraphics[width=6in]{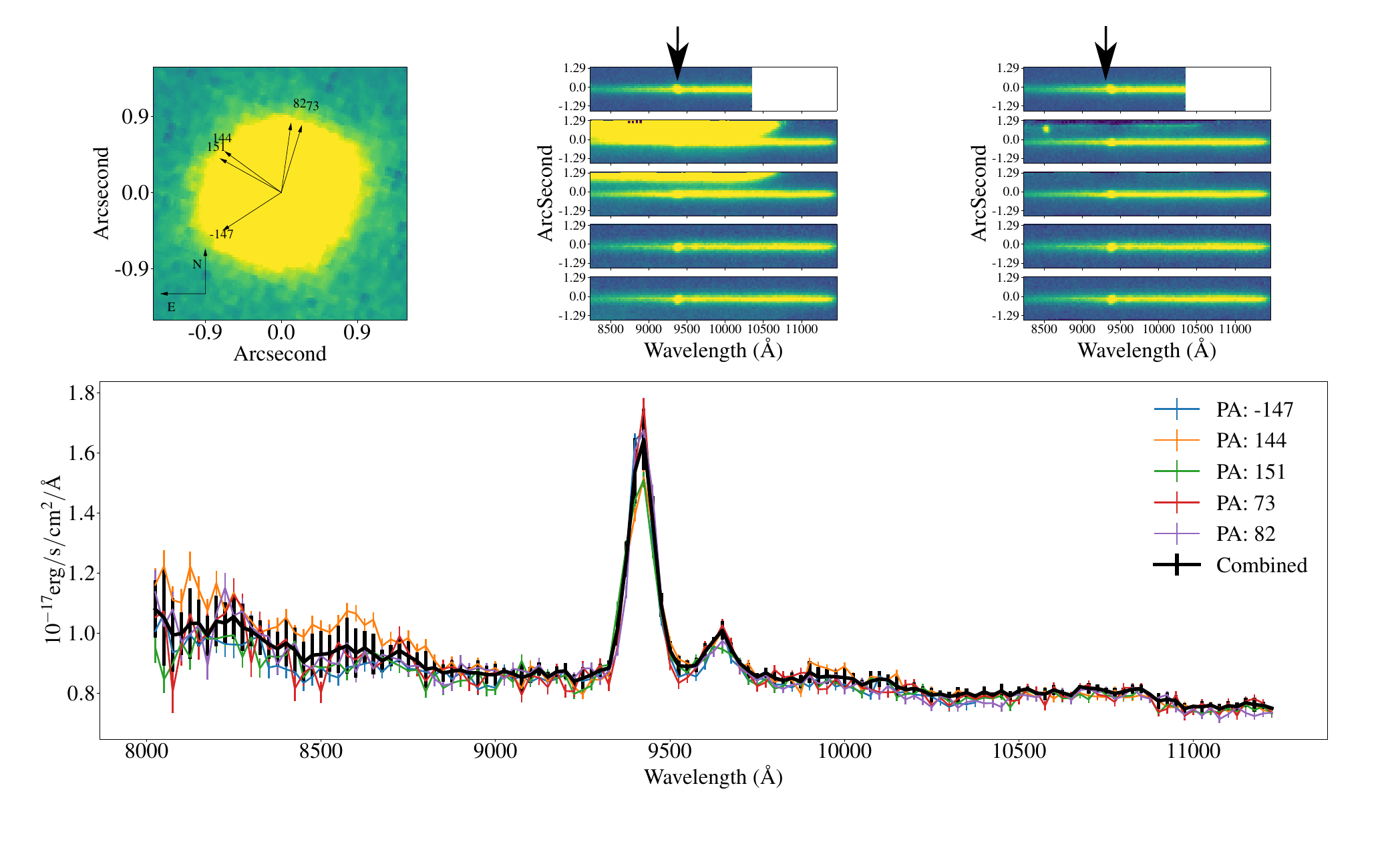}

\caption{A $m_{F105W}=20.23$\ mag source in the GS1 field at a redshift of $z=0.43$\ with prominent H$\alpha$\ and an [SII] lines ($\mathcal{N}=304.$). The H$\alpha$\ line flux is ${\rm 5\times10^{-16}\ erg\ s^{-1}\ cm^{-2}}$. The top left panel shows the F105W image and the orientation of the 5 PAs used to observed this source. The middle top panels shows the 2D rectified spectra. The top right panel shows the contamination-corrected 2D rectified spectra with black arrows pointing to the emission lines. The bottom panel shows the individual 1D spectra (color) as well as the combined spectrum (black).\  lines.\label{FIGS_GS1_2378}}
\end{figure*}

\begin{figure*}
\center
\includegraphics[width=6in]{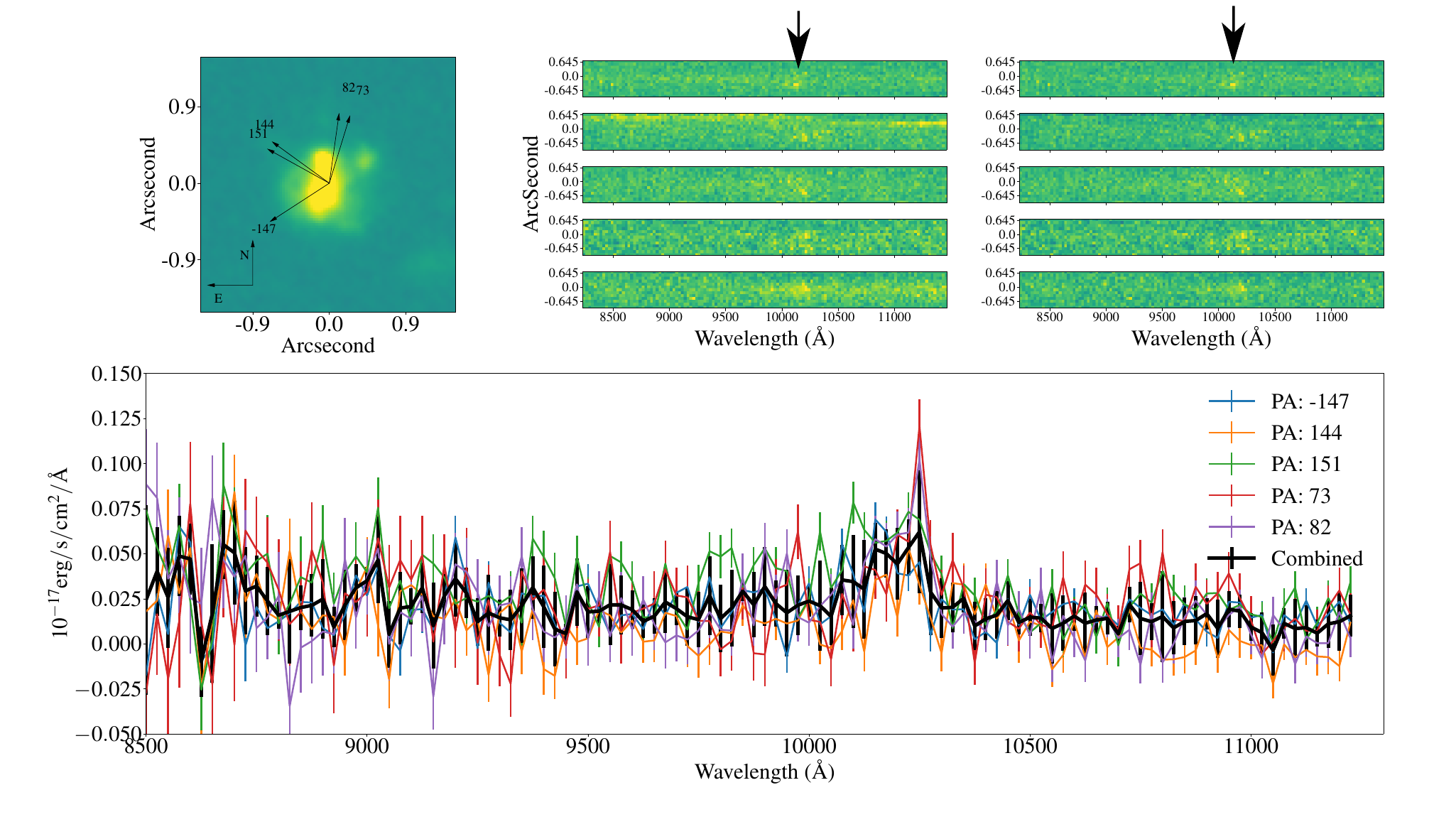}

\caption{A $m_{F105W}=24.1$\ mag source in the GS1 field at a redshift of $z=1.03$\ with and extended [OIII] emission line ($\mathcal{N}=6.95$). The flux of this emission line is ${\rm 3\times10^{-17}\ erg\ s^{-1}\ cm^{-2}}$. The top left panel shows the F105W image and the orientation of the 5 PAs used to observed this source. The middle top panels shows the 2D rectified spectra. The top right panel shows the contamination-corrected 2D rectified spectra with black arrows pointing to the emission line. The bottom panel shows the individual 1D spectra (color) as well as the combined spectrum (black).\  lines.\label{FIGS_GS1_970}}
\end{figure*}

We show two examples of extracted FIGS spectra in Figures \ref{FIGS_GS1_2378} and \ref{FIGS_GS1_970}, in addition to Figure \ref{FIGS_GN1_1496_flx} where we showed the spectra of a large  galaxy at $z_{phot}\approx0.42$. The first shows a relatively bright galaxy with a prominent H$\alpha$\ line, as well as an [SII] line. This Figure shows the level of consistency one can expect between observations taken at different PA, even though the amount of contamination varies greatly. The second example shows a much fainter galaxy with unresolved [OIII] emission, which can be seen to have structure in the 2D spectra, and correlates with the clumpy morphology of this galaxy.
An example of a very faint high redshift FIGS galaxy at a redshift of 7.51 with a ${\rm 1.06 \times 10^{-17} erg\ s^{-1}\ cm^{-2}}$\ Lyman-$\alpha$\ emission and possibly NV emission, suggestive that this source might contain an Active Galaxy Nucleus, can be found in Figure 1 of \citet{Tilvi16}.

\section{Conclusion} \label{sec:conc}
We have described the survey design, methodology, data reduction, and data analysis of FIGS, a deep WFC3/IR slitless spectroscopic survey using the G102 grism filter. We have extracted spectra of 8645 individual sources, 1296 of which are of objects brighter than $m_{F105W}=26.5$\ mag that were observed to the full survey depth of 40 orbits per field. 

The reduced and calibrated FIGS spectra will be made available via the MAST data archive and will include the single-PA 2D data stamps, the single PA extracted spectra using optimal extraction, and finally, the combined 5-PA versions of the 1D spectra. We anticipate these to become available in the latter part of the year 2017.

We expect these deep, multiple PA slitless observations to pave the way for new deep slitless observations from new space based missions:
First, most instruments with the \textit{James 
Webb Space Telescope} (JWST) will be grism-capable, in fact several will have multiple grism 
elements that disperse the light in orthogonal directions to mitigate contamination from overlapping 
spectral traces.  Second, the Wide-Field Infrared Survey Telescope (WFIRST), which is expected to 
survey $\gtrsim\!2\,000$~deg$^2$ and 
collect $\gtrsim\!10^7$~redshifts at $1\!\lesssim\!z\!\lesssim\!3$  as a means of testing the current cosmological model \citep{wfirst}.

\section{Acknowledgments}
We thank the referee for their careful reading of the manuscript and for their insightful suggestions.

Based on observations made with the NASA/ESA Hubble Space Telescope, obtained [from the Data Archive] at the Space Telescope Science Institute, which is operated by the Association of Universities for Research in Astronomy, Inc., under NASA contract NAS 5-26555. These observations are associated with program \#13779 

Support for program \#13779 was provided by NASA through a grant from the Space Telescope Science Institute, which is operated by the Association of Universities for Research in Astronomy, Inc., under NASA contract NAS 5-26555.\\
AC acknowledges the grants ASI n.I/023/12/0 "Attività relative alla fase B2/C
per la missione Euclid" and and MIUR PRIN 2015 "Cosmology and Fundamental 
Physics: illuminating the Dark Universe with Euclid"\\
LC is supported by DFF – 4090-00079.

This work was funded by NASA JWST Interdisciplinary Scientist grants to
RAW NAG5-12460 and NNX14AN10G from GSFC.

\end{document}